\documentclass[nofootinbib,floats,aps,superscriptaddress,preprint]{revtex4}
\usepackage{amssymb,amsmath,amsbsy}
\usepackage{mathrsfs}

\def\bed{\begin{description}}
\def\eed{\end{description}}
\def\iinv{$I$-invariant}
\def\mathbold#1{\boldsymbol{#1}}

\def\to{\rightarrow}

\def\nn{\nonumber}
\def\alv{|\lam_5|}

\def\ycomb{(Y_{11}-Y_{22})}

\def\bea{\begin{eqnarray}}
\def\eea{\end{eqnarray}}

\def\Tr{{\rm Tr}}

\def\tmz{\mathcal{T}}
\def\tm1{{\mathcal{T}}^{-1}}
\def\ttmz{\widetilde{\mathcal{T}}}
\def\ttm1{\widetilde{{\mathcal{T}}}^{-1}}
\def\abar{{\bar a}}
\def\bbar{{\bar b}}
\def\cbar{{\bar c}}
\def\dbar{{\bar d}}
\def\ebar{{\bar e}}
\def\fbar{{\bar f}}
\def\gbar{{\bar g}}
\def\hb{{\bar h}}

\def\lama{\lambda_1}
\def\lamb{\lambda_2}
\def\lamc{\lambda_3}
\def\lamd{\lambda_4}
\def\lame{\lambda_5}
\def\lamf{\lambda_6}
\def\lamg{\lambda_7}
\def\lames{\lambda_5^*}
\def\lamfs{\lambda_6^*}
\def\lamgs{\lambda_7^*}
\def\Lama{\lambda_1^\prime}
\def\Lamb{\lambda_2^\prime}
\def\Lamc{\lambda_3^\prime}
\def\Lamd{\lambda_4^\prime}
\def\Lame{\lambda_5^\prime}
\def\Lamf{\lambda_6^\prime}
\def\Lamg{\lambda_7^\prime}
\def\Maa{m_{11}^{\prime\,2}}
\def\Mbb{m_{22}^{\prime\,2}}
\def\Mab{m_{12}^{\prime\,2}}
\def\lam{\lambda}

\def\ben{\begin{enumerate}}
\def\een{\end{enumerate}}
\def\beq{\begin{equation}}
\def\eeq{\end{equation}}
\def\beqa{\begin{eqnarray}}
\def\eeqa{\end{eqnarray}}

\def\ifmath#1{\relax\ifmmode #1\else $#1$\fi}
\def\lsim{\mathrel{\raise.3ex\hbox{$<$\kern-.75em\lower1ex\hbox{$\sim$}}}}
\def\gsim{\mathrel{\raise.3ex\hbox{$>$\kern-.75em\lower1ex\hbox{$\sim$}}}}
\def\sect#1{section~\ref{#1}}
\def\eq#1{eq.~(\ref{#1})}
\def\Ref#1{ref.~\cite{#1}}
\def\Refs#1#2{refs.~\cite{#1} and \cite{#2}}

\def\eqs#1#2{eqs.~(\ref{#1}) and (\ref{#2})}

\def\eqthree#1#2#3{eqs.~(\ref{#1}), (\ref{#2}) and (\ref{#3})}
\def\eqst#1#2{eqs.~(\ref{#1})--(\ref{#2})}
\def\Eq#1{Eq.~(\ref{#1})}

\def\Eqs#1#2{Eqs.~(\ref{#1}) and (\ref{#2})}

\def\vev#1{\langle #1 \rangle}

\def\ket#1{\left| #1\right\rangle}

\def\sthreet{s_{3\theta}}

\def\cthreet{c_{3\theta}}

\def\lamtil{\lam\ls{345}}

\def\phm{\phantom{-}}

\def\beq{\begin{equation}}
\def\eeq{\end{equation}}
\def\ifmath#1{\relax\ifmmode #1\else $#1$\fi}

\def\st  {s_{\theta}}
\def\ct  {c_{\theta}}
\def\stwot  {s_{2\theta}}
\def\ctwot  {c_{2\theta}}

\def\lamtil{\lam_{345}}

\def\mw{m_W}

\def\ls#1{\ifmath{_{\lower1.5pt\hbox{$\scriptstyle #1$}}}}
\def\lss#1{\ifmath{^{\,\lower2.5pt\hbox{$\scriptstyle #1$}}}}

\def\half{\ifmath{{\textstyle{\frac{1}{2}}}}}

\def\quarter{\ifmath{{\textstyle{\frac{1}{4}}}}}

\def\ie{{\it i.e.}}

\def\etc{{\it etc.}}

\def\npb#1#2#3{{\sl Nucl. Phys. }{\bf B#1}, #3 (#2)}
\def\aop#1#2#3{{\sl Ann.~of~Phys.~}{\bf #1}, #3 (#2)}

\newcounter{mylistcount}
\renewcommand{\Re}{{\rm Re}}
\renewcommand{\Im}{{\rm Im}}

\begin{document}
%
\preprint{
\vbox{\vspace*{2cm}
     \hbox{SCIPP-04/16}
     \hbox{UCD-2004-29}
      \hbox{hep-ph/0506227}
}}
\vspace*{4cm}

\title{Conditions for CP-Violation in the General \\
Two-Higgs-Doublet Model }

\author{John F. Gunion}
\affiliation{ Davis Institute for High Energy Physics \\
University of California, Davis, CA 95616, U.S.A.}

\author{Howard E. Haber}
\affiliation{Santa Cruz Institute for Particle Physics  \\
   University of California, Santa Cruz, CA 95064, U.S.A.}

\begin{abstract}
  
The most general Higgs potential of the two-Higgs-doublet model
(2HDM) contains three squared-mass parameters and seven quartic
self-coupling parameters.  Among these, one squared-mass parameter
and three quartic coupling parameters are potentially complex.  The
Higgs potential explicitly violates CP symmetry if and only if no
choice of basis exists in the two-dimensional Higgs ``flavor'' space
in which all the Higgs potential parameters are real.  We exhibit
four independent potentially complex invariant (basis-independent)
combinations of mass and coupling parameters and show that the
reality of all four invariants provides the necessary and sufficient
conditions for an explicitly CP-conserving 2HDM scalar potential.
Additional potentially complex invariants can be constructed that
depend on the Higgs field vacuum expectation values (vevs).  We
demonstrate how these can be used together with the vev-independent
invariants to distinguish between explicit and spontaneous
CP-violation in the Higgs sector.

\end{abstract}

\maketitle

\section{Introduction}  \label{sec:intro}

The Standard Model (SM) posits the existence of a single complex
hypercharge-one Higgs doublet~\cite{hhg}.
Due to the form of the Higgs potential,
one component of this Higgs scalar acquires a vacuum expectation value
and the SU(2)$\times$U(1) electroweak symmetry is spontaneously broken
to U(1)$_{\rm EM}$.  Hermiticity requires that the parameters of
the SM Higgs potential are real.  Consequently, the resulting bosonic
sector of the electroweak theory is CP-conserving.  CP-violation
enters through the Yukawa couplings of the Higgs field to fermions.
Although there are many potentially complex parameters in the Higgs
couplings to three generations of quarks and leptons, one can redefine
the fermion fields (to absorb unphysical phases).  The end result is one
CP-violating parameter---the Cabibbo-Kobayashi-Maskawa angle~\cite{ckm}.

There are a number of motivations for considering extended Higgs
sectors.  For example, the minimal supersymmetric extension of the
Standard Model requires two complex Higgs doublets~\cite{susyhiggs}.
In this paper, we
consider the most general two-Higgs-doublet extension of the Standard
Model.  This model possesses two identical complex, hypercharge-one
Higgs doublets.  In contrast to the Standard Model, the scalar Higgs
potential of the two-Higgs-doublet model (2HDM) contains potentially
complex parameters~\cite{branco}. 
Consequently, the purely bosonic sector can exhibit
explicit CP-violation (prior to the introduction of the
fermions and the attendant complex Higgs-fermion Yukawa couplings).
However as above, not all complex phases are
physical. In this paper, we exhibit the necessary and sufficient conditions
for an explicitly CP-conserving 2HDM scalar potential.

The procedure for determining whether the Higgs potential explicitly
violates CP is in principle straightforward.  The Higgs potential
parameters are initially defined with respect to two identical
Higgs fields $\Phi_1$ and $\Phi_2$.  However, one can always choose to
change the basis (in the two-dimensional Higgs ``flavor'' space) by
defining two new  (orthonormal)
linear combinations of $\Phi_1$ and $\Phi_2$.  In this
new basis, all the Higgs potential parameters are modified.  
The Higgs potential is
explicitly CP-violating if and only if 
no choice of basis exists in which all the Higgs potential parameters
are simultaneously real.\footnote{We find it convenient and illuminating
to give an explicit proof of this oft-stated result 
in Appendix A.}
If (at least) one basis choice exists in which all Higgs potential
parameters are real, then the Higgs potential is \textit{explicitly}
CP-conserving.  Henceforth, we designate any such basis as
a \textit{real basis}. 
CP-violation in the scalar sector might still
arise if the scalar field vacuum 
is not time-reversal invariant.
In this case, CP is spontaneously
broken~\cite{Lee:1973iz}.

Given an arbitrary Higgs potential, it may not be possible
to determine by inspection whether a real basis exists.
Since there exist four
potentially complex parameters in the Higgs potential, one must in
general solve a set of four non-linear equations (requiring
that these four parameters are real
in some specific basis to be determined).  Thus, we propose another
technique for answering the question of whether a special basis exists
in which all Higgs potential parameters are real.  Our procedure makes
use of the technology introduced in \Ref{davidson} based on
invariant combinations of Higgs potential parameters.  By definition,
these invariants are basis-independent quantities; \textit{i.e.},
they do not depend on the
initial basis choice for $\Phi_1$ and $\Phi_2$.
We then search for potentially complex invariants.

Four potentially complex (basis-independent) invariants govern
the CP-property of the 2HDM scalar potential.
If any one of these four invariants possesses a
non-zero imaginary part, then the 2HDM scalar potential is explicitly
CP-violating.  CP is explicitly conserved if and
only if all four invariants are real.  In the latter case, a real
basis must exist (even though an explicit form for the transformation
that produces such a basis is not determined).  Two of the
invariants were found by diagrammatic techniques in \Ref{davidson}.
Recently, three of the four invariants were also employed
in~\cite{brs}. 
Other earlier simple (basis-dependent) conditions
proposed for the existence of explicit CP-violation in the Higgs
potential~\cite{Grzadkowski:1999ye,ghdecoupling} turn out to be
sufficient but \textit{not} necessary for an explicitly CP-conserving
Higgs potential.

Finally, we note that in the discussion above, we have not
addressed the question of the minimization of the Higgs potential.
This determines the vacuum expectation values (vevs) of the two Higgs
fields,\footnote{We shall
always assume that the Higgs potential parameters are chosen such that
the scalar minimum of interest preserves U(1)$_{\rm EM}$.} 
which are basis-dependent quantities.  
The two vevs can in general be complex, although one can absorb
these complex phases by phase redefinitions of the individual scalar
fields~\cite{Ginzburg:2004vp}.  As shown in Appendix F,
the Higgs sector is fully CP-conserving
if and only if there exists a real basis in which the Higgs vacuum
expectation values are simultaneously real.  The latter can be
established by examining three additional invariants (initially
introduced in \Ref{lavoura}) that depend explicitly on the vevs.

In Section II, the basis-independent formalism for the 2HDM
developed in \Ref{davidson} is reviewed.  In Section III, we exhibit
a set of four
independent potentially complex invariants constructed from the Higgs
sector parameters.  We then prove that the imaginary parts of
these four invariants vanish if and only if the 2HDM scalar potential
explicitly conserves the CP symmetry.  The proof of this theorem
relies on a number of important lemmas that are proved in Appendices C
and D.  The power of this theorem is
demonstrated by exhibiting three simple 2HDM models with complex
parameters that are CP-conserving.
In Section~IV we provide some insight into how the set of four complex
invariants was discovered by surveying all potentially complex $n$th-order
invariants for $n\leq 6$.   The manifest reality of all invariants of
order three or less is demonstrated explicitly in Appendix E.  Thus, one must
search for invariants of order $n\geq 4$ to find candidates that
are potentially complex.
From the results of our survey, we deduce a number of
general features of the potentially complex invariants of arbitrary order.
To determine whether an explicitly CP-conserving Higgs potential
exhibits spontaneous CP-violation, one must additionally consider
basis-independent quantities, initially
introduced in \Ref{lavoura}, that depend on the Higgs vevs.
Finally, a brief discussion of future directions
and concluding remarks are given in Section~VI.

\section{The Higgs potential of the Two Higgs doublet model}
\label{sec:two}

Consider the most general
two-Higgs doublet extension of the Standard Model 
(2HDM) \cite{hhg,Diaz:2002tp}.
Let $\Phi_1$ and
$\Phi_2$ denote two complex $Y=1$, SU(2)$\ls{L}$ doublet scalar fields.
The most general SU(2)$_L\times$U(1)$_Y$ invariant scalar potential is given
by (see, \textit{e.g.}, \Ref{Haber:1993an})
\beqa  \label{pot}
\mathcal{V}&=& m_{11}^2\Phi_1^\dagger\Phi_1+m_{22}^2\Phi_2^\dagger\Phi_2
-[m_{12}^2\Phi_1^\dagger\Phi_2+{\rm h.c.}]\nonumber\\[8pt]
&&\quad +\half\lambda_1(\Phi_1^\dagger\Phi_1)^2
+\half\lambda_2(\Phi_2^\dagger\Phi_2)^2
+\lambda_3(\Phi_1^\dagger\Phi_1)(\Phi_2^\dagger\Phi_2)
+\lambda_4(\Phi_1^\dagger\Phi_2)(\Phi_2^\dagger\Phi_1)
\nonumber\\[8pt]
&&\quad +\left\{\half\lambda_5(\Phi_1^\dagger\Phi_2)^2
+\big[\lambda_6(\Phi_1^\dagger\Phi_1)
+\lambda_7(\Phi_2^\dagger\Phi_2)\big]
\Phi_1^\dagger\Phi_2+{\rm h.c.}\right\}\,,
\eeqa
where $m_{11}^2$, $m_{22}^2$, and $\lam_1,\cdots,\lam_4$ are real parameters
and $m_{12}^2$, $\lambda_5$, $\lambda_6$ and $\lambda_7$ are
potentially complex parameters.  We assume that the
parameters of the scalar potential are chosen such that
the minimum of the scalar potential respects the
U(1)$\ls{\rm EM}$ gauge symmetry.  Then, the scalar field
vacuum expectations values are of the form
\beq \label{potmin}
\langle \Phi_1 \rangle={\frac{1}{\sqrt{2}}} \left(
\begin{array}{c} 0\\ v_1\end{array}\right), \qquad \langle
\Phi_2\rangle=
{\frac{1}{\sqrt{2}}}\left(\begin{array}{c}0\\ v_2\, e^{i\xi}
\end{array}\right)\,,
\eeq
where $v_1$ and $v_2$ are real and non-negative, $0\leq |\xi|\leq\pi$, and
\beq \label{v246}
v^2\equiv v_1^2+v_2^2={\frac{4\mw^2}{g^2}}=(246~{\rm GeV})^2\,.
\eeq
In writing \eq{potmin},
we have used a global U(1)$_Y$ hypercharge rotation to eliminate the phase
of $v_1$.

Since the scalar doublets $\Phi_1$ and $\Phi_2$
have identical SU(2)$\times$U(1) quantum numbers, one
is free to define two orthonormal linear combinations of the original
scalar fields.  The parameters appearing in \eq{pot} depend on a
particular \textit{basis choice} of the two scalar fields.
Relative to an initial (generic) basis choice, the scalar
fields in the new basis are given by
$\Phi^\prime=U\Phi$~\cite{davidson},
where $U$ is a U(2) matrix:\footnote{This U(2) transformation
has also been recently exploited in \Ref{Ginzburg:2004vp}.}
\beq \label{u2}
U=e^{i\psi}\begin{pmatrix} \cos\theta & e^{-i\xi}\sin\theta \\
-e^{i\chi}\sin\theta & e^{i(\chi-\xi)}\cos\theta \end{pmatrix}\,.
\eeq
Note that the
phase $\psi$ has no effect on the scalar potential parameters, since
this corresponds to a global hypercharge rotation.

With respect to the new $\Phi'$-basis, the scalar potential takes on the same
form given in \eq{pot} but with new coefficients
$m_{ij}^{\prime\, 2}$ and $\lambda^\prime_j$.  
 For the general U(2)
transformation of \eq{u2} with $\Phi^\prime=U\Phi$, the scalar
potential parameters ($m_{ij}^{\prime\,2}$, $\lambda_i^\prime$) are related
to the original parameters ($m_{ij}^2$, $\lambda_i$)  by:
\beqa
\Maa&=&m_{11}^2\ct^2+m_{22}^2\st^2-\Re(m_{12}^2 e^{i\xi})\stwot\,,
\label{maa}\\
\Mbb &=&m_{11}^2\st^2+m_{22}^2\ct^2+\Re(m_{12}^2 e^{i\xi})\stwot\,,
\label{mbb}\\
\Mab e^{i\chi}&=&
\half(m_{11}^2-m_{22}^2)\stwot+\Re(m_{12}^2 e^{i\xi})\ctwot
+i\,\Im(m_{12}^2 e^{i\xi})\,.\label{mab}
\eeqa
and
\beqa
\!\!\!\!\!\!\!\!\!\!\!\!\!\!\Lama&=&
\lam_1\ct^4+\lam_2\st^4+\half\lamtil\stwot^2
+2\stwot\left[\ct^2\Re(\lam_6 e^{i\xi})+\st^2\Re(\lam_7e^{i\xi})\right]\,,
\label{Lam1def}  \\
\!\!\!\!\!\!\!\!\!\!\!\!\!\!\Lamb &=&
\lam_1\st^4+\lam_2\ct^4+\half\lamtil\stwot^2
-2\stwot\left[\st^2\Re(\lam_6 e^{i\xi})+\ct^2\Re(\lam_7e^{i\xi})\right]\,,
\label{Lam2def}      \\
\!\!\!\!\!\!\!\!\!\!\!\!\!\!\Lamc &=&
\quarter\stwot^2\left[\lam_1+\lam_2-2\lamtil\right]
+\lam_3-\stwot\ctwot\Re[(\lam_6-\lam_7)e^{i\xi}]\,,
\label{Lam3def}      \\
\!\!\!\!\!\!\!\!\!\!\!\!\!\!\Lamd &=&
\quarter\stwot^2\left[\lam_1+\lam_2-2\lamtil\right]
+\lam_4-\stwot\ctwot\Re[(\lam_6-\lam_7)e^{i\xi}]\,,
\label{Lam4def}      \\
\!\!\!\!\!\!\!\!\!\!\!\!\!\!\Lame  e^{2i\chi} &=&
\quarter\stwot^2\left[\lam_1+\lam_2-2\lamtil\right]+\Re(\lam_5 e^{2i\xi})
+i\ctwot\Im(\lam_5 e^{2i\xi})-\stwot\ctwot\Re[(\lam_6-\lam_7)e^{i\xi}]
\nonumber \\
&&\qquad\qquad\qquad\qquad\qquad\qquad\,\,\,\,
-i\stwot\Im[(\lam_6-\lam_7)e^{i\xi})]\,,
\label{Lam5def}      \\
\!\!\!\!\!\!\!\!\!\!\!\!\!\!\Lamf e^{i\chi}&=&
-\half\stwot\left[\lam_1\ct^2
-\lam_2\st^2-\lamtil\ctwot-i\Im(\lam_5 e^{2i\xi})\right]
+\ct\cthreet\Re(\lam_6 e^{i\xi})+\st\sthreet\Re(\lam_7 e^{i\xi})\nonumber \\
&&\qquad\qquad\qquad\qquad\qquad\qquad\,\,\,\,
+i\ct^2\Im(\lam_6 e^{i\xi})+i\st^2\Im(\lam_7 e^{i\xi})\,,
\label{Lam6def}      \\
\!\!\!\!\!\!\!\!\!\!\!\!\!\!\Lamg e^{i\chi} &=&
-\half s_{2\theta}\left[\lam_1\st^2-\lam_2\ct^2+
\lamtil c_{2\theta}+i\Im(\lam_5 e^{2i\xi})\right]
+\st\sthreet\Re(\lam_6 e^{i\xi})+\ct\cthreet\Re(\lam_7e^{i\xi})\nonumber \\
&&\qquad\qquad\qquad\qquad\qquad\qquad\,\,\,\,
+i\st^2\Im(\lam_6 e^{i\xi})+i\ct^2\Im(\lam_7e^{i\xi})\,,
\label{Lam7def}
\eeqa
where
\beq \label{lamtildef}
\lamtil\equiv\lam_3+\lam_4+\Re(\lam_5 e^{2i\xi})\,.
\eeq
These equations exhibit the following features.
If $m_{11}^2=m_{22}^2$ and $m_{12}^2=0$ in some basis
then these two conditions are true in all bases. Likewise,
if $\lam_1=\lam_2$ and $\lam_7=-\lam_6$
in some basis then these latter two conditions are true in all bases.

We noted previously that the parameters $m_{12}^2$, $\lam_5$, $\lam_6$
and $\lam_7$ are potentially complex.  We now pose the following
question: does there exist a so-called \textit{real basis} in which
all the scalar potential parameters are real?  In general, the
existence of a real basis cannot be ascertained by inspection.
In particular, starting
from an arbitrary basis, it may be quite difficult to determine whether
or not there is a choice of $\theta,\chi,\xi$ above such that
all the primed parameters are real. However, in this paper
we will show, using the
basis-independent techniques described in \Ref{davidson}, that there is a
straightforward procedure for determining whether a real basis exists.
To accomplish this goal, we
write the scalar Higgs potential of the 2HDM following
\Refs{branco}{davidson}:
\beq \label{vcovar}
\mathcal{V}=Y_{a\bbar}\Phi_\abar^\dagger\Phi_b
+\half Z_{a\bbar c\dbar}(\Phi_\abar^\dagger\Phi_b)
(\Phi_\cbar^\dagger\Phi_d)\,,
\eeq
where the indices $a$, $\bbar$, $c$ and $\dbar$
run over the two-dimensional Higgs
flavor space and
\beq \label{zsym1}
Z_{a\bbar c\dbar}=Z_{c\dbar a\bbar}\,.
\eeq
Hermiticity of $\mathcal{V}$ implies that
\beq
Y_{a \bbar}= (Y_{b \abar})^\ast\,,\qquad\qquad
Z_{a\bbar c\dbar}= (Z_{b\abar d\cbar})^\ast\,. \label{zsym2}
\eeq

Under a global U(2) transformation, $\Phi_a\to U_{a\bbar}\Phi_b$
(and $\Phi_\abar^\dagger\to \Phi_\bbar^\dagger U^\dagger_{b\abar}$),
where $U^\dagger_{b\abar}U_{a\cbar}=\delta_{b\cbar}$, and
the tensors $Y$ and $Z$ transform covariantly: $Y_{a\bbar}\to U_{a\cbar}
Y_{c\dbar}U^\dagger_{d\bbar}$
and $Z_{a\bbar c\dbar}\to U_{a\ebar}U^\dagger_{f\bbar}U_{c\gbar}
U^\dagger_{h\dbar} Z_{e\fbar g\hb}$.  The use of barred indices is
convenient for keeping track of which indices transform with $U$ and
which transform with $U^\dagger$.  We also introduce the
U(2)-invariant tensor $\delta_{a\bbar}$, which can be used
to contract indices.  In this notation, one can only contract an
unbarred index against a barred index.  For example,
\beq \label{zabdef}
Z^{(1)}_{a\dbar}\equiv \delta_{b\cbar}Z_{a\bbar c\dbar}=Z_{a\bbar b\dbar}\,,
\qquad\qquad
Z^{(2)}_{c\dbar}\equiv \delta_{b\abar}Z_{a\bbar c\dbar}=Z_{a\abar c\dbar}\,.
\eeq
With respect to the $\Phi$-basis of the unprimed scalar fields, we have:
\beqa \label{ynum}
&& Y_{11}=m_{11}^2\,,\qquad\qquad \phantom{Y_{12}}\,\,\,
Y_{12}=-m_{12}^2 \,,\nonumber \\
&& Y_{21}=-(m_{12}^2)^\ast\,,\qquad\qquad Y_{22}=m_{22}^2\,,
\eeqa
and
\beqa \label{znum}
&& Z_{1111}=\lam_1\,,\qquad\qquad \,\,\phantom{Z_{2222}=}
Z_{2222}=\lam_2\,,\nonumber\\
&& Z_{1122}=Z_{2211}=\lam_3\,,\qquad\qquad
Z_{1221}=Z_{2112}=\lam_4\,,\nonumber \\
&& Z_{1212}=\lam_5\,,\qquad\qquad \,\,\phantom{Z_{2222}=}
Z_{2121}=\lam_5^\ast\,,\nonumber\\
&& Z_{1112}=Z_{1211}=\lam_6\,,\qquad\qquad
Z_{1121}=Z_{2111}=\lam_6^\ast\,,\nonumber \\
&& Z_{2212}=Z_{1222}=\lam_7\,,\qquad\qquad
Z_{2221}=Z_{2122}=\lam_7^\ast\,.
\eeqa
For ease of notation, we have omitted the bars from the barred
indices in \eqs{ynum}{znum}.  Since the
tensors $Y_{a\bbar}$ and $Z_{a\bbar c\dbar}$ exhibit tensorial
properties with respect to global U(2) rotations in the Higgs flavor
space, one can easily construct invariants with respect to the U(2) by
forming U(2)-scalar quantities.

In \sect{sec:three}, we shall argue
that the scalar potential is CP-conserving if and only if 
a real basis exists.  In this case, \textit{all}
possible U(2)-invariant scalars are manifestly real.   Conversely, if
the scalar potential explicitly violates CP, then there must exist at
least one manifestly complex U(2)-scalar invariant.  We shall exhibit
the simplest set of independent
potentially complex U(2)-scalar invariants that
can be employed to test for explicit CP-invariance or non-invariance
of the 2HDM scalar potential.

\section{Complex invariants and the conditions for a CP-conserving
2HDM scalar potential}
\label{sec:three}

Given an arbitrary 2HDM Higgs potential, we have already noted that
the scalar potential possesses a number of potentially complex
parameters.  We would like to determine in general
whether this scalar potential is explicitly CP-violating or
CP-conserving.  The answer to this question is governed by a simple
theorem:

{\underline{\bf Theorem 1:}}  The Higgs potential is explicitly
CP-conserving if and only if a basis exists in which all Higgs potential
parameters are real.  Otherwise, CP is explicitly violated.

Although Theorem 1 is well-known and often stated in the literature,
its proof is usually given under the assumption that a convenient
basis has been chosen in which the CP transformation laws of the
scalar fields assume a particularly simple form~\cite{branco}.
In Appendix A, we provide
a general proof of Theorem 1 that does not make any assumption about the
initial choice of scalar field basis.  As already noted, it may be
difficult to determine
whether a basis exists in which all Higgs potential parameters
are real.  Thus, we would like to reformulate Theorem 1 in a
basis-independent language.  That is, we propose to express the
conditions for an explicitly CP-violating (or conserving) Higgs potential
in terms of basis-independent invariants.

Before presenting the basis-independent version of Theorem 1,
let us first enumerate the number of independent
CP-violating phases that exist among
the scalar potential parameters of the 2HDM.  In
\eq{pot}, we have noted four potentially complex parameters:
$Y_{12}\equiv -m_{12}^2$, $\lambda_5$, $\lambda_6$ and $\lambda_7$.
Naively, it appears that there are three independent CP-violating
phases, since one can always perform a phase rotation on one
of the Higgs fields to render one of the complex parameters real.
However, this conclusion is not correct, since
one can utilize a larger SU(2) global
symmetry to absorb additional phases.\footnote{As previously noted, a U(1)
hypercharge global rotation leaves all the scalar parameters
unchanged; that is, the angle $\psi$ in \eq{u2} has no effect.
If one chooses $\psi=\half(\xi-\chi)$, then the matrix $U$ given in
\eq{u2} is an SU(2) matrix.}
An SU(2) global rotation is parameterized by one angle and two
phases.  This can be used to remove one real parameter and two phases
from the initial ten real parameters and four phases that make up the
scalar potential parameters.  Thus, ultimately, the number of physical
parameters of the scalar potential must be given by nine real
parameters and two phases.  Equivalently, there can only be two
independent complex parameters among the physical parameters that
describe the scalar potential.

This result can be derived in a very simple and direct fashion as
follows~\cite{davidson}.  Consider the 
explicit forms of $Z^{(1)}$ and $Z^{(2)}$
defined in \eq{zabdef}:
\beq \label{zonetwo}
Z^{(1)}=\begin{pmatrix}\lama+\lamd\quad &
\lamf+\lamg \\ \lamf^*+\lamg^* \quad & \lamb+\lamd\end{pmatrix}
\,,\qquad\qquad
Z^{(2)}=\begin{pmatrix}\lama+\lamc \quad& \lamf+\lamg \\
\lamf^*+\lamg^* \quad & \lamb+\lamc\end{pmatrix}\,.
\eeq
Note that $Z^{(1)}$ and $Z^{(2)}$ are hermitian matrices that
commute so that they can be simultaneously diagonalized by
a unitary matrix.  It therefore follows that there
exists a basis in which $Z^{(1)}$ and $Z^{(2)}$ are simultaneously
diagonal; that is, $\lamg=-\lamf$.  Once this basis
is established, it is clear that the phase of $\lamf$ and $\lamg$ can
be removed by a U(1) phase rotation of $\Phi_2$.  Thus, a basis can always
be found in which only two parameters $Y_{12}$ and $\lame$ are
complex.  Moreover, the total number of independent real parameters is
nine (since in a basis where $\lamg=-\lamf$, only one of these two parameters is an
independent degree of freedom).  This matches
the counting of parameters given in the previous paragraph.

Based on this parameter counting, one is tempted to conclude that there
should be only two independent potentially complex invariants.
Nevertheless, this intuition is misleading.  The correct statement is
summarized by the following theorem.

{\underline{\bf Theorem 2:}} The necessary and sufficient conditions
for an explicitly CP-conserving 2HDM scalar
potential consist of the (simultaneous) vanishing of
the imaginary parts of four potentially complex invariants:
\beqa
I_{Y3Z} & \equiv &
\Im(Z^{(1)}_{a\cbar}Z^{(1)}_{e\bbar}Z_{b\ebar c\dbar}Y_{d\abar})
\,, \label{y3zdef}\\
I_{2Y2Z} & \equiv & \Im(Y_{a\bbar}Y_{c\dbar}Z_{b\abar d\fbar}Z^{(1)}_{f\cbar})
\,,\label{2y2zdef}\\
I_{6Z} &\equiv &
\Im(Z_{a \bar{b} c \bar{d} } Z^{(1)}_{b \bar{f}} Z^{(1)}_{d \bar{h}}
Z_{f\bar{a} j \bar{k}} Z_{k \bar{j} m \bar{n}} Z_{n \bar{m} h \bar{c}})\,,
\label{6zdef}\\
I_{3Y3Z}&\equiv &\Im(Z_{a\bar cb\bar
  d}Z_{c\bar ed\bar g}Z_{e\bar h f\bar q}Y_{g\bar a}Y_{h\bar b}Y_{q\bar f}
)\,.
\label{3y3zdef}
\eeqa
Henceforth, the imaginary parts of potentially complex invariants
shall be referred to as \textit{\iinv s}.

The case of 
$\lambda_1=\lambda_2$ and
$\lambda_7=-\lambda_6$ is a special isolated point in the scalar
potential parameter space.  In particular, when  $\lam_1=\lam_2$
and $\lam_7=-\lam_6$, the matrices
$Z^{(1)}$ and $Z^{(2)}$ are both proportional to
the unit matrix.  Thus, if both equalities $\lam_1=\lam_2$
and $\lam_7=-\lam_6$ are true in one basis, then
they must also be true in \textit{all} bases [as previously noted below
\eq{Lam1def}].  Thus, Theorem~2 breaks up into two
distinct cases:
\bed
\item{(i)} For the isolated point
$\lambda_1=\lambda_2$ \textit{and} $\lambda_7=-\lambda_6$,
$I_{Y3Z}=I_{2Y2Z}=I_{6Z}=0$ is automatic
[see \eqst{whythreezee}{sixzee}].  In this case,
the necessary and sufficient condition
for an explicitly CP-conserving 2HDM scalar
potential is simply given by $I_{3Y3Z}=0$.

\item{(ii)} Away from the special isolated point of case (i), only three
of the \iinv s need be considered.  Specifically, at
\textit{any} other point of the parameter space, the necessary and
sufficient conditions for an explicitly CP-conserving 2HDM scalar
potential are given by\footnote{If \eq{theoremb} is satisfied in case (ii),
then it follows that $I_{3Y3Z}=0$.  Thus, the latter
is not needed as a separate requirement.
}
\beq
I_{Y3Z}= I_{2Y2Z}=I_{6Z}=0\,.
\label{theoremb}
\eeq

\eed

It is trivial to prove that the above conditions are necessary for
explicit CP-conservation.  If any of the above \iinv s
[\eqst{y3zdef}{3y3zdef}] are non-zero, then we can immediately conclude
that no basis exists in which all scalar potential parameters are
real.   Thus, by Theorem~1, the scalar potential would be CP-violating.
The proof that the conditions of
Theorem~2 are sufficient for explicit CP-conservation will now be given,
with further details provided in  Appendices C and D.

First, we must prove that all four \iinv s listed in \eqst{y3zdef}{3y3zdef}
are required in the formulation of Theorem~2.  This may be
accomplished by exhibiting
four different models in which only one of the four \iinv s
is  non-zero. In Sections IV.A and IV.C, we give
explicit forms for these four \iinv s in a generic basis [see
eqs.~(\ref{yzzz}), (\ref{yyzz}), (\ref{sixz}) and (\ref{i3y3zform}),
respectively].  However, as already noted below \eq{zonetwo}, 
it is always possible to choose a basis
in which $\lambda_7=-\lambda_6$.  This basis is not unique,
since further basis transformations can be performed while maintaining
$\lambda_7=-\lambda_6$.  In any such basis, three of the \iinv s take
particularly simple forms:
\beqa
I_{Y3Z}&=&-(\lama-\lamb)^2\,\Im(Y_{12}\lamf^*)\,,\label{whythreezee}\\
I_{2Y2Z}&=&(\lama-\lamb)\left[\Im(Y_{12}^2\lame^*)+\ycomb
\Im(Y_{12}\lamf^*)\right]\,,\label{twowhytwozee}\\
I_{6Z}&=&-(\lama-\lamb)^3\,\Im(\lamf^2\lame^*)\,.\label{sixzee}
\eeqa
The expression for $I_{3Y3Z}$ in this basis is more complicated:
\beqa
&&
I_{3Y3Z}=2\,\Im(Y_{12}^3\lam_6(\lam_5^*)^2)-4\,\Im(Y_{12}^3(\lam_6^*)^3)
+[\ycomb^2-6|Y_{12}|^2]\ycomb \Im(\lam_6^2\lam_5^*)\nn\\[4pt]
&&\qquad +\left[(\lam_1-\lam_3-\lam_4)(\lam_2-\lam_3-\lam_4)
+2|\lam_6|^2-\alv^2\right]\ycomb\Im(Y_{12}^2\lam_5^*)\nn\\[4pt]
&&\qquad+\Bigl\{(\lam_1-\lam_2)^2 Y_{11}Y_{22}  +(4|\lam_6|^2-2\alv^2)\left[\ycomb^2-|Y_{12}|^2\right]
\Bigr\}\Im(Y_{12}\lam_6^*)\nn\\[4pt]
&&\qquad -(\lam_1+\lam_2-2\lam_3-2\lam_4)\Bigl\{\ycomb
\Im(Y_{12}^2(\lam_6^*)^2)-\Im(Y_{12}^3\lam_5^*\lam_6^*)\nn\\
&&\qquad\qquad\qquad\qquad\qquad\qquad\qquad \qquad
+\left[\ycomb^2-|Y_{12}|^2\right]
\Im(Y_{12}\lam_6\lam_5^*)\Bigr\}\,.
\label{3y3zl6eq-l7}
\eeqa
Working in the
$\lambda_7=-\lambda_6$ basis, we consider the four models:
\begin{enumerate}
\item
$Y_{a\bbar}=0$ and $\lam_1\neq\lam_2$\,;
\item
$\lambda_6=0$\,, $\lam_1\neq\lam_2$ and $Y_{11}=Y_{22}$\,;
\item
$\lambda_5=0$\,, $\lam_1\neq\lam_2$\,, $Y_{11}=Y_{22}=0$
and Re($Y_{12}\lam_6^*)=0$\,;
\item
$\lam_1=\lam_2$\,.
\end{enumerate}
Then, in model 1, $I_{Y3Z}=I_{2Y2Z}=I_{3Y3Z}=0$ whereas $I_{6Z}$ is
potentially non-zero.
In model~2, $I_{Y3Z}=I_{6Z}=I_{3Y3Z}=0$ whereas
$I_{2Y2Z}$ is potentially non-zero.
In model~3, $I_{2Y2Z}=I_{6Z}=I_{3Y3Z}=0$
whereas $I_{Y3Z}$ is potentially non-zero.\footnote{Note that if
$\lam_5=0$ and $Y_{11}=Y_{22}$, then
$I_{3Y3Z}=\left\{(\lam_1-\lam_2)^2 Y_{11}Y_{22}-16
\,[\Re(Y_{12}\lam_6^*)]^2\right\}\, \Im(Y_{12}\lam_6^*)$.}
Finally, in model 4,
$I_{Y3Z}=I_{6Z}=I_{2Y2Z}=0$ whereas $I_{3Y3Z}$ is potentially non-zero.
Thus, we have exhibited four separate models in which CP is violated
explicitly, and in each case only one of the four \iinv s is non-zero.
This illustrates that all four \iinv s are
needed to test whether the Higgs potential explicitly conserves or
violates CP.

The requirement of four \iinv s
in the formulation of Theorem 2 seems to be
in conflict with our previous observation that the number of physical
parameters of the 2HDM includes only \textit{two} phases [see
discussion surrounding \eq{zonetwo}].   However, one can show that for
any particular model, at most two  \iinv s need
be considered.  To verify this assertion, we first transform to a basis
in which $\lambda_7=-\lambda_6$ and where $\lambda_6$ (and therefore
$\lambda_7$) are real.\footnote{This is always possible as shown below
\eq{zonetwo}.}
Then there are a number of cases to consider.  (i) If
$\lambda_1=\lambda_2$, then $I_{3Y3Z}=0$ implies that the Higgs sector
is explicitly CP-conserving.
(ii) If $\lambda_1\neq\lambda_2$ and $Y_{12}$, $\lambda_5$ and
$\lambda_6$ are non-vanishing, then $I_{Y3Z}=I_{6Z}=0$ implies
that the Higgs
sector is explicitly CP-conserving.  (iii) If $\lambda_1\neq\lambda_2$
and two of the quantities $Y_{12}$, $\lambda_5$ and
$\lambda_6$ are non-zero while the third vanishes, then only one \iinv\
need be considered.  Specifically, for $\lam_5=0$ [$\lam_6=0$],
$I_{Y3Z}=0$ [$I_{2Y2Z}=0$] guarantees a CP-conserving Higgs sector, whereas
for $Y_{12}=0$, $I_{6Z}=0$ guarantees a CP-conserving Higgs sector.  
Thus, we have shown that
it is sufficient to examine at most two \iinv s to determine whether
all four \iinv s
[\eqst{y3zdef}{3y3zdef}] simultaneously vanish.\footnote{Of course, to take
advantage of this observation
in practice, one must be able to take the original model and
transform to a basis where $\lambda_7=-\lambda_6$ is real.  In
general, this may be difficult (and require a numerical computation).
Thus, in order to test for explicit CP-violation,
it is often simpler to directly evaluate all four \iinv s in the
original basis.}

To complete the proof of Theorem 2, we must show that if
the four \iinv s given by
\eqst{y3zdef}{3y3zdef} vanish, then one can find a
basis where all Higgs potential parameters are real.
The proof is most easily carried out by first transforming to a basis
in which $\lambda_7=-\lambda_6$ and where $\lambda_6$ (and therefore
$\lambda_7$) are real.  In this basis, the cases of $\lambda_1=\lambda_2$
and $\lambda_1\neq\lambda_2$
must be treated separately.
First, we consider the case where $\lambda_1\neq\lambda_2$.
If $\lambda_7=-\lambda_6\neq 0$, then $I_{6Z}=0$ [\eq{sixzee}]
implies that $\lam_5$ is also real in this basis, and
$I_{Y3Z}=0$ [\eq{whythreezee}] implies that $Y_{12}$ is real.
We have therefore achieved a basis in which all
scalar potential parameters are real.  If $\lam_6=\lam_7=0$, then one can
perform a phase rotation on one of the scalar fields so that $\lam_5$
is real, with $Y_{12}$ potentially complex. In this new basis, if
$\lam_5\neq 0$ then $I_{2Y2Z}=0$ implies that $Y_{12}$ is either
real or purely imaginary.  In the latter case, \eqst{Lam5def}{Lam7def} show
that a U(2) transformation [see \eq{u2}] 
with parameters $\xi=\pi/2$,
$\sin 2\theta=0$ and $\chi=0$ yields a basis in which
$\lam_6'=-\lam_7'=0$, and both $\lam_5'=-\lam_5$ and $Y_{12}'$ are real.
Finally, if
$\lam_5=\lam_6=\lam_7=0$, then one can absorb any phase of
$Y_{12}$ into a phase redefinition of one of the scalar fields.

Next, we consider the case where $\lambda_1=\lambda_2$ in a basis
where $\lam_7= -\lam_6$.  In this case, it is always possible to make a
further change of basis so that $\lam_5$, $\lam_6$ and $\lam_7$ are
real (this assertion is Lemma 2, which is proved in
Appendix C).\footnote{In Appendix C, Lemma 3 demonstrates why the
condition of $\lambda_1=\lambda_2$ is crucial to the proof of Lemma 2.}
In this latter basis where $Y_{12}$ is potentially complex but all other scalar
potential parameters are real, \eq{3y3zl6eq-l7} yields the following
form for the only potentially non-vanishing  invariant $I_{3Y3Z}$:
\beqa \label{3y3zspecial}
&&I_{3Y3Z}= 2\,{\rm Im}~Y_{12}
\left[\lam_5^2+\lam_5 (\lam_1-\lam_3-\lam_4) - 2\lam_6^2\right]\nonumber \\
&&\quad\times \left[4\lam_6\,({\rm Re}~Y_{12})^2-
(\lam_3+\lam_4+\lam_5-\lam_1)(Y_{11}-Y_{22})\,{\rm Re}~Y_{12}
 -\lam_6 (Y_{11}-Y_{22})^2\right]\,.
\eeqa
Then,
$I_{3Y3Z}=0$ implies that one of the following
three conditions must be true in a basis where all the $\lambda_i$ are
real: (i) $Y_{12}$ is real; (ii) the quantity
$\lam_5^2+\lam_5 (\lam_1-\lam_3-\lam_4) - 2\lam_6^2=0$; or (iii) the
quantity $4\lam_6\,({\rm Re}~Y_{12})^2-
(\lam_3+\lam_4+\lam_5-\lam_1)(Y_{11}-Y_{22})\,{\rm Re}~Y_{12} -\lam_6
(Y_{11}-Y_{22})^2=0$.  In Appendix D, we prove Lemma 4 which
demonstrates that if $Y_{12}$
is complex and either condition (ii) or condition~(iii) holds, then it
is possible to find a basis in which $Y_{12}$ is real, while
maintaining the reality of $\lam_5$, $\lam_6$ and $\lam_7$.  Hence it
follows that if $I_{3Y3Z}=0$, then there exists a basis in which all
2HDM scalar potential parameters are real.\footnote{The
U(2) rotation required to go to this basis is explicitly
constructed in Appendix D.}
The proof of Theorem 2 is now complete.

It is instructive to
compare the results of Theorem 2 to one of the basis-dependent
conditions that has been proposed in the literature.
In a generic basis,
a sufficient set of conditions for an explicitly CP-conserving
2HDM scalar potential is:
\beq \label{relative}
{\rm Im}~(Y_{12}^2\lambda_5^*)={\rm Im}~(Y_{12}\lambda_6^*)
={\rm Im}~(Y_{12}\lambda_7^*)
={\rm Im}~(\lambda_5^*\lambda_6^2)=
{\rm Im}~(\lambda_5^* \lambda_7^2)={\rm Im}~(\lambda_6^* \lambda_7)=0\,,
\eeq
where $Y_{12}\equiv -m_{12}^2$.
Clearly, if \eq{relative} is satisfied, then a simple phase rotation of one
of the scalar fields easily produces a basis in which all the scalar
potential parameters are real.   However, \eq{relative} is not
necessary for CP-conservation.  In particular, the following statement
is generally false:
``the Higgs potential is explicitly CP-violating if one or more of the
quantities listed in \eq{relative} are non-vanishing.''
This is most easily demonstrated by the following exercise.  Start
with a model in which all potentially complex scalar potential
parameters are real.  Then, change the basis with a generic U(2)
transformation [\eq{u2}].  In a typical case, the resulting parameters
$Y_{12}'$, $\lam_5'$, $\lam_6'$, and $\lam_7'$ in the new basis are
complex, and
one or more of the quantities listed in \eq{relative} are non-vanishing.
Thus, \eq{relative} is not a necessary condition for an explicitly
CP-conserving Higgs potential.\footnote{An example that
illustrates the same point is a model in which $\lambda_1=\lambda_2$
and $\lambda_7=-\lambda_6$. Lemma 2 of Appendix C implies that we can
transform to a basis in which all the $\lambda_i$ are real.  Nevertheless,
in this basis, \eq{3y3zspecial} implies that it is possible to have
an explicitly CP-conserving model with $I_{3Y3Z}=0$ and $\Im~Y_{12}\neq 0$.}

Despite the relative simplicity of
the forms for $I_{Y3Z}$, $I_{2Y2Z}$, $I_{6Z}$ and $I_{3Y3Z}$ in the
$\lam_7=-\lam_6$ basis, realistic models rarely conform to this
particular basis choice.   The power of the basis-independent
formulation of Theorem~2 thus becomes evident when considering models
where the transformation from the generic basis to the
$\lam_7=-\lam_6$ basis is not particularly simple.
Fortunately, we possess expressions for
these \iinv s in a generic basis [see
eqs.~(\ref{yzzz}), (\ref{yyzz}), (\ref{sixz}) and (\ref{i3y3zform})],
so there is no compelling need to explicitly perform this change of basis.
For purposes of illustration, let us consider three special models.  In
model (i),
\beq
\lama=\lamb\,,\quad \lamf=\lamg\quad\,{\rm and}\,\quad
Y_{11}=Y_{22}\,, \label{cpa}
\eeq
where $Y_{12}$, $\lam_5$ and $\lam_6$ have arbitrary phases.  In model (ii),
\beq
\lama+\lamb=2(\lamc+\lamd)\,,\quad
\lame=0\,\quad {\rm and}\quad\, \lamf=\lamg\,,\label{cpb}
\eeq
where $Y_{12}$ and $\lam_6$ have arbitrary phases.
In model (iii),
\beq  \label{cpc}
\lama=\lamb\,,\quad \lamf=\lamg^*\,,\quad Y_{11}=Y_{22}\quad
{\rm and}\quad Y_{12}\,,\lambda_5~{\rm real}
\,,
\eeq
where $\lambda_6$ has an arbitrary phase.
Model~(iii) arises by imposing on the Higgs potential
a discrete permutation symmetry that
interchanges $\Phi_1$ and $\Phi_2$~\cite{rebelo}.

In the three models above, we have used
eqs.~(\ref{yzzz}), (\ref{yyzz}), (\ref{sixz}) and (\ref{i3y3zform})
in the generic basis to verify that
$I_{Y3Z}=I_{2Y2Z}=I_{6Z}=I_{3Y3Z}=0$.  Thus models (i), (ii) and (iii)
are explicitly CP-conserving.  These three models also provide
examples of explicitly CP-conserving 2HDM potentials where
\eq{relative} is \textit{not} satisfied.  Nevertheless, having
verified that all the \iinv s vanish, one is assured of the
existence of some basis choice for each model for which all Higgs potential
parameters are real.

Here, we provide one explicit example in the case
of model (iii).  Starting from the generic basis specified in \eq{cpc},
we perform a U(2) transformation [\eq{u2}] with
$\theta=\pi/4$ and $\xi=0$.  Then, \eqthree{mab}{Lam6def}{Lam7def}
yield $m_{12}^{\prime\,2}=\lam'_6=\lam'_7=0$, while \eq{Lam5def} implies
that:
\beq \label{rebelotransform}
\lam'_5 e^{2i\chi}=\half(\lam_1-\lam_3-\lam_4+\lam_5)
-2i\Im~\lam_6\,.
\eeq
It is now a simple matter to adjust $\chi$ so that $\lam'_5$ is real.
Thus, we have exhibited the U(2) transformation that produces the
``real basis''
of model (iii) in which all scalar potential parameters are real.
Applying this U(2)
transformation to the fields, it is easy to check that
the resulting real basis exhibits a discrete symmetry
$\Phi'_1\to\Phi_1'$, $\Phi'_2\to -\Phi_2'$.  Models that respect the latter
discrete symmetry are manifestly CP-invariant since $\lam'_5$ is the
only potentially complex parameter, whose phase can be rotated away by
an appropriate phase rotation of $\Phi'_2\to e^{i\chi}\Phi'_2$.

\section{A survey of complex invariants}
\label{sec:four}

In general, it is possible to construct an $n$th
order invariant quantity for any integer value of $n$, where $n$
is the total number of $Y$'s and $Z$'s that appears in the
invariant. The vast majority of such invariants are manifestly real.
In this section, we focus on those invariants that are potentially
complex.

The necessary and sufficient 
conditions for CP-conservation have been presented in
Theorem~2 and depend on only four potentially complex invariants given
by \eqst{y3zdef}{3y3zdef}.  However, new potentially complex $n$th order
invariants arise at every order (for $n>4$) that cannot
be expressed in terms of lower-order invariants.  Nevertheless, Theorem~2
guarantees that if the \iinv s of \eqst{y3zdef}{3y3zdef}
vanish, then the imaginary parts of \textit{all} potentially
complex invariants must vanish.  In particular, we have explicitly
verified the following statements:
\vskip 0.05in

\begin{enumerate}
\setcounter{enumi}{\value{mylistcount}}
\item
All invariants (of arbitrary order)
that are either independent of $Z$ or linear in $Z$ are
manifestly real.
\item
All invariants of cubic order or less are manifestly real.
\item
Any
quartic (\textit{i.e.}, fourth-order)
\iinv\ is a real linear
combination of $I_{Y3Z}$ and $I_{2Y2Z}$.
\item
Any fourth or higher-order
\iinv\ that is quadratic in $Z$ is proportional to $I_{2Y2Z}$.
\item
Any fifth-order \iinv\
vanishes if $I_{Y3Z}=I_{2Y2Z}=0$.
\item
Any sixth-order \iinv\ that is independent of $Y$
is proportional to $I_{6Z}$.  Moreover, if $Y_{a\bbar}=0$
then any \iinv\
of arbitrary order vanishes if $I_{6Z}=0$.
\item
Any sixth order \iinv\ that is both cubic in $Y$
and $Z$ respectively is a real linear combination of 
$I_{3Y3Z}$ and lower-order invariants that vanish if $I_{Y3Z}=I_{2Y2Z}=0$.
\item
Any sixth order \iinv\ that is either linear or quadratic in $Y$
vanishes if $I_{Y3Z}=I_{2Y2Z}=0$.
\setcounter{mylistcount}{\value{enumi}}
\end{enumerate}

\noindent
Finally, we reiterate that:

\begin{enumerate}
\setcounter{enumi}{\value{mylistcount}}
\item
Any \iinv\ of
arbitrary order vanishes if $I_{Y3Z}=I_{2Y2Z}=I_{6Z}=I_{3Y3Z}=0$.
\setcounter{mylistcount}{\value{enumi}}
\end{enumerate}

\noindent
This last result is
a consequence of Theorem~2.  The explicit verification of
statements 1--8 is based on a systematic study of potentially complex
U(2)-invariant scalars made up of the tensors $Y_{a\bbar}$ and
$Z_{a\bbar c\dbar}$.  This study gives us further confidence that the
ultimate conclusion given by statement 9 above is correct.

We begin this study by noting that for
$n=1$, the only invariants are $\Tr~{Y}$, $\Tr~Z^{(1)}$ and
$\Tr~Z^{(2)}$, all of which are manifestly real.  For $n=2$, the 
possible quadratic
invariants include the products of the first order invariants and
$\Tr~(Y^2)$, $\Tr~(YZ^{(1)})$, $\Tr~(YZ^{(2)})$, $\Tr~(Z^{(i)}Z^{(j)})$
[$(i,j)=(1,1), (1,2), (2,2)$], $\Tr~Z^{(31)}
\equiv Z_{a\bbar c\dbar}Z_{b\abar d\cbar}$
and $\Tr~Z^{(32)}\equiv Z_{a\bbar c\dbar}Z_{d\abar b\cbar}$,
where $Z^{(31)}$ and $Z^{(32)}$ are introduced in
\eq{zthreej}.\footnote{Note that the
determinants of $Y$, $Z^{(1)}$ and $Z^{(2)}$ are also 
quadratic invariants, but
these can be expressed in terms of invariants already given above due to the
identity $\det M\equiv \half[(\Tr~M)^2-\Tr~(M^2)]$ which is satisfied by any
$2\times 2$ matrix.}  By inspection, all such quadratic invariants are
manifestly real.  Turning to the cubic invariants, the enumeration of
all possible cases becomes significantly more complex.  Nevertheless,
as shown in Appendix E,
it is still possible to verify by hand that all cubic invariants are
manifestly real.  
Thus, in order
to find a potentially complex invariant, one must examine invariants
of fourth order and higher.  At this point, an explicit hand
calculation becomes infeasible, and we must employ a computer
algebra program such as Mathematica to assist in the analysis.
For example, consider all possible invariants that are independent of
$Y_{a\bbar}$ (such invariants will be called $Z$-invariants).
One can use Mathematica to evaluate the imaginary part of each
invariant by explicitly considering invariants which
consist of $n$-fold products of $Z$'s. These invariants are of
the form:
\beq
Z_{a_1\bar b_1 c_1\bar d_1} Z_{a_2\bar b_2 c_2\bar d_2}\cdots
Z_{a_n\bar b_n c_n\bar d_n}\,,
\eeq
where one chooses the indices $\left\{b_1,d_1,b_2,d_2,\ldots,b_n,d_n\right\}$
to be a particular permutation of
$\left\{a_1,c_1,a_2,c_2,\ldots,a_n,c_n\right\}$, and then sums over
the repeated indices as usual.  By considering all possible permutations,
one generates all $(2n)!$ possible invariants
(many of which are trivially related to others in the complete list
of invariants).
One can automate the computation with a Mathematica program and
compute the imaginary part of all $(2n)!$ invariants subject to
the constraints of computer time. The procedure can be generalized
to include some number of $Y_{a\bbar}$.
In particular, it is easy to show (without computer assistance)
that all invariants that are independent of $Z$
(such invariants will be called $Y$-invariants) are manifestly real,
due to the hermiticity property of $Y_{a\bbar}$.

\medskip
\subsection{Fourth-order potentially complex invariants}

Among the quartic invariants, we first construct all possible
quartic $Z$-invariants.
By an explicit Mathematica
computation, we were able to show that all $8!=40,320$
quartic $Z$-invariants are manifestly real.

We next search for potentially complex quartic invariants that are linear in
$Y$.  We display one potentially non-zero \iinv\ below:
\beqa \label{yzzz}
I_{Y3Z}&\equiv &
\Im(Z^{(1)}_{a\cbar}Z^{(1)}_{e\bbar}Z_{b\ebar c\dbar}Y_{d\abar})\nonumber \\
&=& 2(|\lamf|^2-|\lamg|^2)\Im[Y_{12}(\lamfs+\lamgs)]
+(\lama-\lamb)\bigl[\Im(Y_{12}\Lambda^*)
-\Im[Y_{12}\lames(\lamf+\lamg)]\bigr]\nonumber
\\ && +\ycomb\bigl[
\Im[\lames(\lamf+\lamg)^2]-(\lama-\lamb)\Im(\lamgs\lamf)\bigr]
\,,
\eeqa
where
\beq \label{Lamdef}
\Lambda\equiv (\lamb-\lamc-\lamd)\lamf+(\lama-\lamc-\lamd)\lamg\,.
\eeq
Using Mathematica, we have evaluated the imaginary part of all
7!= 5,040 possible invariants that are linear in $Y$ and cubic in $Z$.  We
find that the result either vanishes or is equal to $\pm I_{Y3Z}$.

Next, we examine potentially complex quartic
invariants that are quadratic in both $Y$ and $Z$. We
display one potentially non-zero \iinv\ below:
\beqa \label{yyzz}
I_{2Y2Z} &\equiv &\Im(Y_{a\bbar}Y_{c\dbar}Z_{b\abar
  d\fbar}Z^{(1)}_{f\cbar}) \nonumber \\
&=& (\lama-\lamb)\Im(Y_{12}^2\lames)
-\ycomb\left[\Im(Y_{12}\Lambda^*)+\Im(Y_{12}\lam_5^*(\lam_6+\lam_7))\right]
\nonumber\\
&&-\Im[(Y_{12}\lamfs)^2]+\Im[(Y_{12}\lamgs)^2]
+\left[\ycomb^2-2|Y_{12}|^2\right]\Im(\lamgs\lamf)\,.
\eeqa
Moreover, we find as before that the imaginary parts of all such invariants
(there are 6!=720 invariants that are quadratic in both $Y$ and $Z$)
either vanish or are equal to $\pm I_{2Y2Z}$.

It is easy to show that quartic invariants that are cubic in $Y$
(and therefore linear in $Z$) are manifestly real.  In particular, there
are only two such invariants that are not a product of lower order
invariants: $\Tr~(Y^3 Z^{(1)})$ and $\Tr~(Y^3 Z^{(2)})$.  Both these
invariants are manifestly real due to the hermiticity properties of
$Y$, $Z^{(1)}$ and $Z^{(2)}$.

\medskip

\subsection{Fifth order potentially complex invariants}

We begin by constructing all possible
fifth-order $Z$-invariants.
Again, with the help of Mathematica, we found that all
$10!=3,628,800$ $Z$-invariants are manifestly real.
Next, we considered the $Y4Z$-invariants, \textit{i.e.} the
fifth-order invariants that are linear in $Y$.
After computing the imaginary parts of all $9!=362,880$ such
invariants, we found that only one genuinely new potentially complex
invariant emerged.  The corresponding \iinv\ is :
\beqa
I_{Y4Z}&=&\Im[Z^{(2)}_{a\bar b} Z_{b\bar a c \bar d} Z^{(2)}_{d\bar e}
Z_{e\bar c f \bar g } Y_{g\bar f}]\nn\\
&=& -\lambda_4 I_{Y3Z}
+(\lam_1-\lam_2)\Im[Y_{12}(\lam_6^*+\lam_7^*)^2(\lam_6^*-\lam_7^*)]\nn\\
&& +\Im[Y_{12}\lam_5^*(\lam_6^2\lam_7^*-\lam_7^2\lam_6^*)]
+\Im[Y_{12}\lam_5(\lam_6^{*\,2}-\lam_7^{*\,2})(\lam_6^*+\lam_7^*)]\nn\\
&&
+\half\left(-(\lam_1-\lam_2)(\lam_1+\lam_2-2\lam_3-2\lam_4)
+|\lam_7|^2-|\lam_6|^2\right)\Im[Y_{12}\lam_5^*(\lam_6+\lam_7)]\nn\\
&&-\half\left((\lam_1-\lam_2)^2-|\lam_6|^2-|\lam_7|^2\right)
\Im[Y_{12}\lam_5^*(\lam_6-\lam_7)]\nn\\
 &&+\half(\lam_1-\lam_2)(2|\lam_5|^2-|\lam_6|^2-|\lam_7|^2)
\Im[Y_{12}(\lam_6^*+\lam_7^*)]\nn\\
&&+\half(\lam_1-\lam_2)(|\lam_6|^2-|\lam_7|^2)
\Im[Y_{12}(\lam_6^*-\lam_7^*)]\nn\\
&&+\half(Y_{11}-Y_{22})\Bigl[4(|\lam_6|^2-|\lam_7|^2)\Im(\lam_6\lam_7^*)
+(\lam_1-\lam_2)\Im[\lam_5^*(\lam_7^2-\lam_6^2)]\nn\\
&&\quad\quad\quad\quad\quad\quad\quad\quad
+(\lam_1+\lam_2-2\lam_3-2\lam_4)\Im[\lam_5^*(\lam_6+\lam_7)^2]\Bigr]\,,
\label{yzzzz}
\eeqa
where $I_{Y3Z}$ is given by \eq{yzzz}.  In addition, we have
explicitly verified that the imaginary parts of all
potentially complex $Y4Z$-invariants
reduce to a linear combination of $I_{Y4Z}$ and the product of
$I_{Y3Z}$ times a linear combination of $\Tr[Z^{(1)}]$ and
$\Tr[Z^{(2)}]$.

The fact that $I_{Y4Z}$ is a ``new'' \iinv\ means that
one cannot express $I_{Y4Z}$ as a sum of terms, each of
which is the imaginary part of a product of lower-order invariants.
Nevertheless, one can show that if
$I_{Y3Z}=I_{2Y2Z}=0$, then it follows that $I_{Y4Z}=0$.
This is most easily accomplished in the basis where
$\lambda_7=-\lambda_6$.  In this basis, \eq{yzzzz} simplifies
enormously:
\beq \label{yzzzzshort}
I_{Y4Z}=(\lambda_1-\lambda_2)^2\left[\lambda_4\Im(Y_{12}\lambda_6^*)
-\Im(Y_{12}\lambda_6\lambda_5^*)\right]\,.
\eeq
If $Y_{12}=0$ in the $\lambda_7=-\lambda_6$ basis then $I_{Y4Z}=0$.
Alternatively, if $Y_{12}\neq 0$, then we make use of:
\beq \label{imy1265}
\Im(Y_{12}\lambda_6\lambda_5^*)=\frac{1}{|Y_{12}|^2}
\left[\Im(Y_{12}^2\lambda_5^*)\Re(Y_{12}\lambda_6^*)-
\Im(Y_{12}\lambda_6^*)\Re(Y_{12}^2\lambda_5^*)\right]\,.
\eeq
Since $I_{Y3Z}=I_{2Y2Z}=0$ implies that 
either $\lambda_1=\lambda_2$ or $\Im(Y_{12}^2\lambda_5^*)=
\Im(Y_{12}\lambda_6^*)=0$ [see \eqs{whythreezee}{twowhytwozee}],
one can again conclude that
$I_{Y4Z}=0$.  Having proved that the \textit{invariant} 
$I_{Y4Z}$ vanishes in one basis, it immediately follows
that $I_{Y4Z}=0$ in \textit{all} basis choices.

Similarly, we have analyzed the $2Y3Z$-invariants, \textit{i.e.}, the
fifth-order invariants that are
quadratic in $Y$ and cubic in $Z$.  Again, we have computed the
imaginary parts of all $40,320$ such invariants.  We have
explicitly verified that any potentially complex fifth-order invariant
of this type is a linear combination of $I_{Y3Z}$ (with coefficient
proportional to $\Tr~Y$),
$I_{2Y2Z}$
(with coefficient proportional to a linear combination of $\Tr[Z^{(1)}]$
and $\Tr[Z^{(2)}$]) and one new potentially complex invariant form.
A particular choice for the new \iinv\ is:
\beq
I_{2Y3Z}=\Im[Z_{a\bar c b\bar e}Z_{c\bar f d\bar b}Z_{e\bar g f\bar
  h}Y_{g\bar a} Y_{h\bar d}]\,.
\label{i2y3zdef}
\eeq
One could write out the explicit expression for $I_{2Y3Z}$ [as we did
in \eq{yzzzz} for $I_{Y4Z}$].  However, for
our purposes, it is sufficient to give the form of $I_{2Y3Z}$
in the $\lam_7=-\lam_6$ basis:
\bea
I_{2Y3Z}&=&
(\lam_1-\lam_2)\left[4\Im[Y_{12}^2\lam_6^{*\,2}]+2(Y_{11}-Y_{22})
\Im[Y_{12}\lam_5^*\lam_6]-(\lam_1+\lam_2-2\lam_3)
\Im[Y_{12}^2\lam_5^*]\right.\nn\\[5pt]
&&\qquad\qquad \left. -2\lam_4 (Y_{11}-Y_{22})\Im(Y_{12}\lambda_6^*)\right]\,.
\eea
Again, it must be emphasized that $I_{2Y3Z}$ is a ``new'' \iinv\
in the sense that
one cannot express $I_{2Y3Z}$ as a sum of terms, each of
which is the imaginary part of
a product of lower-order invariants.  Nevertheless,
$I_{Y3Z}=I_{2Y2Z}=0$ implies that $I_{2Y3Z}=0$.\footnote{This is
easily verified after
noting that $\Im(Y_{12}^2\lambda_6^{*2})=2\Im(Y_{12}\lambda_6^*)
\Re(Y_{12}\lambda_6^*)$.}

The remaining cases are easily treated.  We explicitly verified that
any fifth-order invariants that is
cubic in $Y$ and quadratic in $Z$ is proportional to $(\Tr~Y) I_{2Y2Z}$.
It is also simple to show that
all fifth-order invariants that are linear in $Z$ are
manifestly real. In particular, the only two inequivalent invariants
of this type that are not
products of lower order invariants are $Z_{a\bbar c
\dbar}Y^2_{b\abar}Y^2_{d\cbar}$ and $Z_{a\bbar c
\dbar}Y^2_{b\cbar}Y^2_{d\abar}$.  By explicit calculation, using the
hermiticity properties of $Y$ and $Z$, it is straightforward to verify
that both these invariants are real.
We have previously noted that all pure
$Y$-invariants are manifestly real.
This completes the proof that all
potentially complex fifth-order invariants are linear
combinations of $I_{Y3Z}$ and $I_{2Y2Z}$ or forms
that vanish when $I_{Y3Z}=I_{2Y2Z}=0$.
That is, the consideration of potentially complex fifth order
invariants does not establish any new independent conditions for CP violation.

\medskip

\subsection{Sixth-order potentially complex invariants}

Two new independent conditions for CP violation arise from the study
of sixth-order potentially complex invariants. We begin by
constructing all possible sixth-order $Z$-invariants. It is here that
we encounter the
first potentially complex $Z$-invariants. One potentially
non-zero \iinv\ is:
\beqa \label{sixz}
I_{6Z}&\equiv &
\Im(Z_{a \bar{b} c \bar{d} } Z^{(1)}_{b \bar{f}} Z^{(1)}_{d \bar{h}}
Z_{f\bar{a} j \bar{k}} Z_{k \bar{j} m \bar{n}} Z_{n \bar{m} h \bar{c}})\nn\\
&=& 2|\lame|^2\Im[(\lamgs\lamf)^2]
-\Im[{\lames}^{\,2}(\lamf-\lamg)(\lamf+\lamg)^3]+(\lama-\lamb)|\lame|^2\Im[\lames(\lamf+\lamg)^2]\nn\\
&&+2\Im(\lamgs\lamf)\biggl[|\lame|^2[|\lamf|^2
+|\lamg|^2-(\lama-\lamb)^2]-2(|\lamf|^2-|\lamg|^2)^2\biggr]\nonumber \\
&& -(\lama-\lamb)\Im\bigl(\lames\Lambda^2)
-2(|\lamf|^2-|\lamg|^2)\Im[\lames\Lambda(\lamf+\lamg)]
\nonumber \\
&&+(\lam_1-\lam_2)\biggl[\Im\left[\Lambda
(\lam_7\lam_6^{*\,2}+\lam_6\lam_7^{*\,2}-|\lam_7|^2\lam_6^*-|\lam_6|^2\lam_7^*)\right]\nn\\
&&
\hspace*{1in}+2\Im\left[\lam_5\left((|\lam_6|^2+|\lam_7|^2)\lam_6^*\lam_7^*
-\lam_7\lam_6^{*\,3}-\lam_6\lam_7^{*\,3}\right)\right]\biggr]
\,,
\label{i6zform}
\eeqa
where $\Lambda$ is defined in \eq{Lamdef}.

Theorem 2 implies that if
$Y_{a\bar b}=0$ and $I_{6Z}=0$, then any $Z$-invariant is
real. Consequently, the imaginary part of any sixth-order $Z$
invariant must be equal to $cI_{6Z}$, for some real constant $c$.
Our proof of Theorem 2 in \sect{sec:three} leaves no doubt as to the
veracity of this conclusion.  Nevertheless, it is instructive to check
this assertion explicitly.  Unfortunately, a
complete survey of all
possible $12!=479,001,600$ sixth-order
complex $Z$-invariants is beyond the capability of our desktop
computers.  However, we were able to examine roughly nine million
sixth-order $Z$-invariants, and in these cases the imaginary part of each
sixth-order $Z$-invariant either vanishes or is equal to $\pm
I_{6Z}$ or $\pm 2 I_{6Z}$.

If $\lam_1\neq\lam_2$ in a basis where $\lam_7=-\lam_6$, then
Theorem~2 implies that $I_{Y3Z}=I_{2Y2Z}=I_{6Z}=0$ is a 
necessary and sufficient
condition for an explicitly CP-conserving 2HDM scalar
potential.  However, the case of $\lam_1=\lam_2$ and $\lam_7=-\lam_6$
(where $I_{Y3Z}=I_{2Y2Z}=I_{6Z}=0$ is automatic) must be treated
separately.  In this latter case, the condition for CP violation
depends on an independent invariant that first arises at sixth order
and is made up of three $Y$ and three $Z$ factors (henceforth denoted
as $3Y3Z$-invariants).

Thus, we have constructed all possible $3Y3Z$-invariants and examined
their imaginary parts.  Of course, some of these will simply be linear
combinations of lower-order invariants already examined.  A complete
survey of the imaginary part of
all possible 9!=362,880 $3Y3Z$-invariants yields one new
independent \iinv.  A representative choice is:
\bea
I_{3Y3Z}&=&\Im(Z_{a\bar cb\bar
  d}Z_{c\bar ed\bar g}Z_{e\bar h f\bar q}Y_{g\bar a}Y_{h\bar b}Y_{q\bar f}
)\nn\\
&=&(Y_{11}-Y_{22})\left[(\lam_1-\lam_3-\lam_4)(\lam_2-\lam_3-\lam_4)
-\alv^2+|\lam_6|^2+|\lam_7|^2\right] \Im[Y_{12}^2\lam_5^*] \nn \\
&&\quad
+[(Y_{11}-Y_{22})^2-|Y_{12}|^2]\,\alv^2\,\Im[Y_{12}(\lam_7^*-\lam_6^*)]
-Y_{11}Y_{22}\,(\lam_1-\lam_2)\,\Im[Y_{12}\Lambda^*]\nn\\
&&\quad
+2\,[(Y_{11}-Y_{22})^2+Y_{11}Y_{22}-|Y_{12}|^2]\left[|\lam_7|^2
\Im(Y_{12}\lam_6^*)-|\lam_6|^2\Im(Y_{12}\lam_7^*)\right]
\nn\\
&&\quad
+2\,Y_{11}Y_{22}\left[|\lam_7|^2\,\Im(Y_{12}\lam_7^*)
-|\lam_6|^2\Im(Y_{12}\lam_6^*)\right]\nn\\
&&\quad
+(\lam_1-\lam_2)Y_{11}Y_{22}\Im[Y_{12}\lam_5^*(\lam_6+\lam_7)]
-[\ycomb^2-|Y_{12}|^2]\Im(Y_{12}\lam_5^*\widetilde\Lambda)\nn\\
&&\quad
-\ycomb\Bigl\{(Y_{11}Y_{22}+|Y_{12}|^2)
\bigl[\Im[\lam_5^*(\lam_6^2+\lam_7^2)]
-(\lam_1-\lam_2)\Im(\lam_6\lam_7^*)\bigr]\nn \\
&&\qquad\qquad
+(Y_{11}^2+Y_{22}^2-4|Y_{12}|^2)\Im[\lam_5^*\lam_6\lam_7]
-(\lam_1+\lam_2-2\lam_3-2\lam_4)\Im[Y_{12}^2\lam_6^*\lam_7^*]\Bigr\}\nn \\
&&\quad
+\Im[Y_{12}^3\lam_5^*\widetilde\Lambda^*]
+2\,\Im[Y_{12}^3\lam_6^*\lam_7^*(\lam_6^*-\lam_7^*)]
+\Im[Y_{12}^3(\lam_5^*)^2(\lam_6-\lam_7)]\,,
\label{i3y3zform}
\eea
where $\Lambda$ is defined in \eq{Lamdef} and
\beq \label{Lambdatildadef}
\widetilde\Lambda\equiv(\lam_2-\lam_3-\lam_4)\lam_6
-(\lam_1-\lam_3-\lam_4)\lam_7\,.
\eeq
We have explicitly verified that the imaginary part of any $3Y3Z$
invariant is a real linear combination of $I_{3Y3Z}$,
$(\Tr~Y) I_{2Y3Z}$, $[\Tr~Y]^2 I_{Y3Z}$, $[\Tr~Y^2] I_{Y3Z}$,
$\Tr~[YZ^{(1)}]I_{2Y2Z}$,
$\Tr~[YZ^{(2)}]I_{2Y2Z}$ and $(\Tr~Y\,\Tr~Z^{(1)})I_{2Y2Z}$.\footnote{%
Note that
$\Tr~Y\Tr~Z^{(2)}=\Tr~Y\Tr~Z^{(1)}-\Tr~[Y(Z^{(1)}-Z^{(2)})]$.}
In a basis where $\lam_7=-\lam_6$, $I_{3Y3Z}$ reduces to the
expression given by \eq{3y3zl6eq-l7}.  Indeed,
$I_{3Y3Z}$ is non-zero in explicitly CP-violating models with
$\lam_7=-\lam_6$ and
$\lam_1=\lam_2$, which confirms that it is a necessary ingredient in
the formulation of Theorem~2.

Among other sixth order invariants, all $6Y$ and $Z5Y$ invariants are
manifestly real.  A $2Z4Y$ invariant is potentially complex, but its
imaginary part must be proportional to some linear combination of
$(\Tr~Y)^2 I_{2Y2Z}$~and  $(\Tr~Y^2) I_{2Y2Z}$.  This leaves two
interesting cases: the $Y5Z$ and $2Y4Z$ invariants, which we now
consider in more detail.

A partial scan of the imaginary part of $10!=3628800$
$2Y4Z$-invariants and $11!=39916800$ $Y5Z$-invariants
has been performed, and our results
yield two genuinely new potentially complex invariants, whose
imaginary parts we 
designate by $I_{2Y4Z}$ and $I_{Y5Z}$, respectively.
The resulting expressions in a generic basis
are quite complicated and not very illuminating.  Hence, here we
provide only the explicit forms in a basis where $\lam_7=-\lam_6$:
\bea
I_{2Y4Z}&=& \Im(Z_{b\bar c}^{(2)} Z_{c\bar e d\bar f} Z_{e\bar q f\bar
  r} Z_{g\bar b h \bar d} Y_{q\bar g} Y_{r\bar h})\nn\\
&=& (\lam_1-\lam_2)\Bigl\{ (\lam_1+\lam_2)\Im(Y_{12}^2\lam_6^{*\,2})
-(\lam_1\lam_2-|\lam_5|^2-2|\lam_6|^2)\Im(Y_{12}^2\lam_5^*)\nn\\
&&\qquad +\left[ 2 (\lam_1 Y_{11}-\lam_2 Y_{22})
-(\lam_3+\lam_4)(Y_{11}-Y_{22})\right]\Im(Y_{12}\lam_6\lam_5^*)\nn\\
&&\qquad
+\left[2|\lam_5|^2(Y_{11}-Y_{22})-(\lam_3+\lam_4)(\lam_1Y_{11}-\lam_2
Y_{22})\right]\Im(Y_{12}\lam_6^*)\nn \\
&&\qquad
-\left[(Y_{11}-Y_{22})^2-2|Y_{12}|^2\right]\,\Im(\lam_6^2\lam_5^*)\Bigr\}\,,
\eea
and
\bea
I_{Y5Z}&=& \Im(Z_{b\bar c}^{(1)} Z_{c\bar b d\bar e} Z_{e\bar d f\bar
  g} Z_{g\bar q}^{(1)} Z_{q\bar f r\bar s} Y_{s\bar r})\nn\\
&=& (\lam_1-\lam_2)^2
\Bigl\{(Y_{11}-Y_{22})
  \Im(\lam_6^2\lam_5^*)-(\lam_1+\lam_2)\,\Im(Y_{12}\lam_6\lam_5^*)\nn\\
&&\qquad\qquad\qquad
+[\lam_4(\lam_1+\lam_2)+\lam_4^2-|\lam_5|^2]\Im(Y_{12}\lam_6^*)\Bigr\}\,.
\eea
If $Y_{12}\neq 0$ in the $\lambda_7=-\lambda_6$ basis, then we can
use:
\beq
\Im(\lambda_6^2\lambda_5^*)=\frac{1}{|Y_{12}|^2}\left[\Re(Y_{12}\lam_6^*)
\,\Im(Y_{12}\lam_6\lam_5^*)-\Re(Y_{12}\lam_6\lam_5^*)\,
\Im(Y_{12}\lam_6^*)\right]
\eeq
along with \eqthree{whythreezee}{twowhytwozee}{imy1265} to conclude
that both $I_{2Y4Z}$ and $I_{Y5Z}$ vanish if
$I_{Y3Z}=I_{2Y2Z}=0$.\footnote{Although
this result is demonstrated in the $\lam_7=-\lam_6$ basis, the
conclusion must hold for all basis choices.}
However, if
$Y_{12}=0$ in the $\lambda_7=-\lam_6$ basis, then \textit{all}
invariants of $n$th order with $n\leq 5$ are real.  In the latter
case, both $I_{2Y4Z}$ and $I_{Y5Z}$ can still be non-vanishing,
which demonstrates that these are
new \iinv s.  Nevertheless, by the same argument as
before, we may conclude that if $I_{Y3Z}=I_{2Y2Z}=I_{6Z}=0$, then
both $I_{2Y4Z}$ and $I_{Y5Z}$ must vanish.  For this reason, $I_{2Y4Z}$
and $I_{Y5Z}$ need not be independently considered
in the formulation of Theorem~2.\footnote{However, it is
not possible to express either $I_{2Y4Z}$ or $I_{Y5Z}$
as a linear combination of $I_{6Z}$
$I_{Y3Z}$ and $I_{2Y2Z}$ with corresponding coefficients that
are invariant quantities.}
In particular, $I_{6Z}$ is included in the
statement of Theorem 2, since (unlike $I_{2Y4Z}$ and $I_{Y5Z}$)
$I_{6Z}$ can be non-zero even when $Y_{a\bar b}=0$.

We have verified\footnote{Our conclusion is based on a partial scan of
about two million invariants.  However, the arguments in the next
sub-section strongly suggest that the following results apply to all
$2Y4Z$ and $Y5Z$ invariants.}
that the imaginary part of any $2Y4Z$ invariant
can be expressed as a real
linear combination of  $I_{2Y4Z}$, $I_{2Y3Z}$, $I_{2Y2Z}$ and
$I_{Y3Z}$.  Likewise,  the imaginary part of any $Y5Z$ invariant
can be expressed as a real
linear combination of  $I_{Y5Z}$, $I_{Y4Z}$ and $I_{Y3Z}$.
In both cases, each of the corresponding coefficients of the linear
combination of terms are real invariant quantities.
As an explicit illustrative example, we have verified:
\beqa
&&\hspace{-0.5in}
{\rm Im}\left[Z^{(2)}_{b\bar{c}}Z_{c\bar{b} d\bar{e}}Z_{e\bar{d}
f\bar{g}}Z_{g\bar{q} h\bar{r}}Y_{q\bar{f}} Y_{r\bar{h}}\right]=
I_{2Y4Z}-\half\left[
\Tr(Z^{(1)}Z^{(2)})-\half[\Tr
Z^{(1)}]^2+Z_{a\bar{c}b\bar{d}}Z_{c\bar{a}d\bar{b}}\right]I_{2Y2Z}
\nn\\[5pt]
&& \qquad\qquad\quad +\quarter\Tr Z^{(1)}I_{2Y3Z}+\half\Tr Y
I_{Y4Z}-\half\left[\Tr(Z^{(1)}Y)
+\half{\rm Tr}~Y~\Tr Z^{(2)}\right]I_{Y3Z}\,.
\eeqa

\medskip

\subsection{General results for $\boldsymbol{n}$th-order
potentially complex invariants}

The analyzes of Sections IV.A and IV.B
permit us to conjecture a number
of results that we expect to hold for complex invariants of arbitrary
order.  These results provide a method for identifying the number of
``new'' potentially complex invariants at any order.  As before, we
define a ``new'' $n$th order \iinv\ to be one that cannot be written as a
sum of terms, each of which is the imaginary part of a product of
known invariants of order $\leq n$.
By this definition, ``new'' \iinv s arise at each order (for $n\geq
4$).  However, as previously stated, if
$I_{Y3Z}=I_{2Y2Z}=I_{6Z}=I_{3Y3Z}=0$, then any new
\iinv\ that arises must also vanish.

Consider an arbitrary $n$th order \iinv\ $I_{pYqZ}$
made up of $p$ factors of $Y$ and $q=n-p$ factors of $Z$.  In a basis where
$\lambda_7=-\lambda_6$, for $p\leq 3$ 
\beq \label{pyqz}
I_{pYqZ}=(\lambda_1-\lambda_2)^{3-p}~\Im~P(Y_{12},\lam_5,\lam_6)\,,
\eeq
where $P$ is a polynomial of its arguments and their complex
conjugates constructed such that each term in the sum
contains $p$ factors of $Y_{a\bar b}$ and $q+p-3$ factors of the $\lambda_i$,
with the constraint that the weight of each term
in the sum is zero.  Here, we define the weight $w$ according to the rules:
$w(Y_{12})=+1$, $w(\lam_5)=+2$, $w(\lam_6)=+1$, $w(x^*)=-w(x)$ for
any $x$ and $w(xy)=w(x)+w(y)$ for any $x$, $y$.\footnote{Formally, the
weight $w=w(x)$ for any scalar potential parameter $x$ is defined such
that $x\to e^{iw\theta}x$ under a redefinition of
one of the scalar fields by $\Phi_1\to e^{i\theta}\Phi_1$.
Of course, $w=0$ for any real scalar potential parameter.}

The polynomial $P$ possesses one additional property of note: it does
not vanish in the limit of $\lam_1=\lam_2$ (assuming that $P\neq 0$ in
general).  That is, the behavior of $I_{pYqZ}$ in the $\lam_1\to\lam_2$
limit is specified explicitly in \eq{pyqz}.
If $p>3$, then $I_{pYqZ}=0$.  For example at sixth order,
$\Tr~(Y^2)I_{2Y2Z}$ is a potentially non-vanishing \iinv\ with
$p=4$, but this does not constitute a new \iinv\ by the above
definition.

\Eq{pyqz} is consistent with all the results of Sections IV.A and IV.B.
It also provides an explanation for the absence of complex invariants
of low order.  For example, if we apply
\eq{pyqz} and attempt to construct $I_{5Z}$, we would need to find a
polynomial $P$ with a non-zero imaginary part that is quadratic
in the $\lam_i$.  No such polynomial exists, and we conclude that
$I_{5Z}=0$.  We can also use \eq{pyqz} to predict the results of
higher order invariants.  For example, all seventh and eighth order
$Z$-invariants must be proportional to $I_{6Z}$ (a result that we have
confirmed by limited scanning).  However, a ``new'' $Z$-invariant arises
at ninth order, which in the $\lam_7=-\lam_6$ basis must 
have an imaginary part that is a linear
combination of $I_{6Z}P_3(\lambda_i)$ and
$(\lam_1-\lam_2)^3\,\Im[(\lam_6^2\lam_5^*)^2]$, where $P_3(\lambda_i)$
is a real cubic polynomial of the $\lambda_i$.
Although this is a new \iinv, it clearly vanishes when $I_{6Z}=0$.

Finally, \eq{pyqz} strongly suggests that there is only one new $2Y4Z$
\iinv\ and one new $Y5Z$ \iinv, since in each case, only one
new term, $\Im(\lam_6^2\lam_5^*)$ arises that did not appear in
lower-order invariants (in the $\lam_7=-\lam_6$
basis).\footnote{Unfortunately, this argument fails to explain the
existence of only one $3Y3Z$ \iinv, a fact that has been confirmed
only by a complete scan over all possible invariants of this type.}

\section{Implications for spontaneous CP-violation}
\label{sec:five}

If a Higgs potential is explicitly CP-conserving, then there exists a
so-called ``real basis'' in which all the Higgs potential
parameters are real.  A theory with an explicitly CP-conserving Higgs sector
may be CP-violating if the vacuum does not respect the CP-symmetry.
In this case, we say that CP is spontaneously broken~\cite{Lee:1973iz}.
To determine whether CP is spontaneously broken, one must check
whether the vacuum is invariant under time reversal.
We assert the following theorem, which is proved in Appendix F:

\underline{\bf Theorem 3:}
Given an explicitly CP-conserving Higgs potential,
the vacuum is time-reversal invariant if and only if
a real basis exists in which
the Higgs vacuum expectation values are real.

Theorem 3 requires one to
verify the existence or nonexistence of a basis with certain
properties.  However, these theorems can be reformulated in a
basis-independent language.  Here, we follow \Ref{lavoura}, and
introduce three U(2)-invariants~\cite{davidson}:
\beqa
-\half v^2 J_1 &\equiv & \widehat v_\abar^\ast Y_{a\bbar} Z^{(1)}_{b\dbar}
\widehat v_d\,, \label{cpxinvariants1}\\
\quarter v^4
J_2&\equiv& \widehat v_\bbar^\ast \widehat v_\cbar^\ast Y_{b\ebar}
Y_{c\fbar} Z_{e\abar f\dbar} \widehat v_a \widehat v_d\,,
 \label{cpxinvariants2}\\
J_3 &\equiv & \widehat  v_\bbar^\ast \widehat v_\cbar^\ast
Z^{(1)}_{b\ebar}Z^{(1)}_{c\fbar}Z_{e\abar f\dbar} \widehat v_a
\widehat v_d\,,
 \label{cpxinvariants3}
\eeqa
where $\vev{\Phi^0_a}\equiv v \widehat v_a/\sqrt{2}$, with $v=246$~GeV
and $\widehat v$ is a unit vector in the complex two-dimensional Higgs
flavor space.  The scalar potential minimum condition is easily
derived from \eq{vcovar}:
\beq \label{potmingeneric}
\widehat v_\abar^\ast\,
[Y_{a\bar b}+\half v^2 Z_{a\bbar c\dbar}\, \widehat v_\cbar^\ast\,
\widehat v_d]=0 \,.
\eeq
Thus, we may eliminate $Y$ in the expressions for $J_1$ and
$J_2$:
\beqa
J_1 &\equiv & \widehat v_\abar^\ast \widehat v_\ebar^\ast
Z_{a\bbar e\fbar} Z^{(1)}_{b\dbar}
\widehat v_d \widehat v_f\,, \label{cpxinvariants1p}\\
J_2&\equiv& \widehat v_\bbar^\ast \widehat v_\cbar^\ast
\widehat v_{\bar g}^\ast \widehat v_{\bar p}^\ast
Z_{b\ebar g\bar h}Z_{c\fbar p \bar r}
Z_{e\abar f\dbar} \widehat v_a \widehat v_d \widehat v_h \widehat v_r\,.
 \label{cpxinvariants2p}
\eeqa
Since $\Im~Y_{12}$ is determined by the scalar potential minimum
conditions in terms of $\Im~\lambda_{5,6,7}$, one is left with
three potentially complex parameters in a basis where
$\widehat v$ is real.  These are in one-to-one correspondence with
$J_1$, $J_2$ and $J_3$.

\underline{\bf Theorem 4:}
Consider the 2HDM scalar potential in some arbitrary basis.  Assume
that the minimum of the scalar potential preserves U(1)$_{\rm EM}$.
Then, the Higgs sector is CP-conserving (\ie, no explicit nor
spontaneous CP-violation is present) if $J_1$, $J_2$ and $J_3$ defined in
\eqst{cpxinvariants1}{cpxinvariants3} are real~\cite{lavoura}.

If the Higgs sector is CP-conserving, then according to Theorem 3
some basis must exist in which the Higgs potential
parameters and the Higgs field vacuum expectation values are
simultaneously real.  But in that case, we may
immediately conclude that the  \textit{invariant} quantities
$J_1$, $J_2$ and $J_3$ must be real.  Conversely,
the reality of $J_1$, $J_2$ and $J_3$ provide
sufficient conditions for a CP-invariant
Higgs sector.  This result is proven in
refs.~\cite{lavoura} and \cite{branco},%
\footnote{In fact,
there are at most two independent
relative phases among $J_1$, $J_2$ and $J_3$.  However,
as shown in \Ref{davidson}, there are
cases where two of the three invariants are real and only one has an
non-vanishing imaginary part,
which shows that one must check all three invariants in order to
determine whether the Higgs sector is CP-invariant.}
and we do not repeat the proof here.

Note that \eqst{cpxinvariants1}{cpxinvariants3} are considerably
simpler than the invariants that govern explicit CP-violation of the
Higgs potential [\eqst{y3zdef}{3y3zdef}].  However, these two sets of
invariants serve different purposes.  To answer the question of
whether the Higgs sector is CP-invariant, one must first choose a
basis and minimize the scalar potential.  Having found $\widehat v_a$,
one may now compute $J_1$, $J_2$ and $J_3$.  If these invariants are
all real, then the Higgs potential is explicitly CP-invariant and there
is no spontaneous CP-violation.  If at least one of the invariants
$J_1$, $J_2$ and $J_3$ is complex, then the Higgs sector is
CP-violating.  However, in this latter case, one must evaluate 
the four \iinv s given in \eqst{y3zdef}{3y3zdef} to
determine whether CP is spontaneously or explicitly broken.
If these four \iinv s all vanish, then CP is
spontaneously broken. If at least one of these is non-zero, then CP is
explicitly broken.  These conclusions are summarized in our final
theorem:

\underline{\bf Theorem 5:}
The necessary and sufficient conditions for spontaneous CP-violation
in the 2HDM are: (i)~$I_{Y3Z}=I_{2Y2Z}=I_{6Z}=I_{3Y3Z}=0$, and
(ii)~at least one of the three invariants $J_1$, $J_2$, and/or $J_3$
possesses a non-vanishing imaginary part.
If (i) is \textit{not} satisfied then (ii) is necessarily
true, and the CP-violation is explicit.  If (ii) is \textit{not} satisfied,
then (i) is necessarily true, and the Higgs sector is CP-conserving.

We provide two simple examples.  First, \Ref{chinese} considers a model
in which $m_{12}^2=\lam_6=\lam_7=0$ and $\lam_5$ is real and
positive.  Minimizing the scalar potential yields a purely
imaginary $v_2/v_1$.  Nevertheless, a simple relative phase
redefinition of the two Higgs fields by $\pi/2$ yields a real basis
with real vacuum expectation values.  (In the new basis, $\lam_5'<0$
and all other Higgs potential parameters are unmodified.)
Hence, this model is CP-conserving.

Second, consider a
Higgs potential that satisfies \eq{cpc}, with $\lambda_6$
real, which was proposed in \Ref{rebelo}.  
That is, all scalar potential parameters of this model are
real, and the Higgs potential is explicitly CP-conserving.
In this case, a minimum of the scalar potential exists where $v_1=v_2$
and the relative phase of the two vevs, $\xi\neq 0$.
That is, we may write $\sqrt{2}\,\widehat v=(e^{-i\xi/2}\,,\,e^{i\xi/2})$.
Nevertheless, \Ref{rebelo} proved that this model is CP-conserving.
We may explicitly verify this assertion by performing a U(2) transformation
given by \eq{u2} with $\psi=\xi/2$, $\chi=\pi/2$ and $\theta=\pi/4$.
We find that $\lambda_5'=-\lambda_5$,
$m_{12}^{\prime\,2}=m_{12}^2\sin\xi$,
$\lambda'_6=\lambda'_7=\lam_6\sin\xi$ are all real
and $\widehat v^{\,\prime}=(1\,,0)$.
Thus, we have established a basis in which all scalar potential
parameters and the vacuum expectation values are simultaneously real.

Of course, the absence of spontaneous CP-breaking in both examples can
also be confirmed by checking that the invariants $J_1$, $J_2$ and
$J_3$ are all real.

\section{Conclusions}
\label{sect:conclude}

The connection between the CP property of a general scalar potential
and the parameters of the potential and vacuum expectation values
of the Higgs fields is governed by two well-known theorems.  
The first, proven here as Theorem 1,
states that the Higgs sector is explicitly CP-conserving if and
only if there exists a \textit{real basis}, that is
choice of basis (in the Higgs ``flavor'' space) 
in which all the scalar potential parameters are real.
The second theorem, proven here as Theorem 3, 
states that the vacuum is CP-invariant,
implying the absence of both explicit and spontaneous CP violation,
if and only if there exists a  real basis in which the Higgs
vacuum expectation values are real. In this paper,
we have established a simple procedure for determining whether or not
a general 2HDM is explicitly CP-conserving by employing 
a set of four potentially complex
basis-independent invariant combinations of the Higgs 
potential parameters.   At least one of these invariants possesses a
non-vanishing imaginary part if and only
if no real basis exists. 

The imaginary parts of the 
four complex basis-independent invariants that govern
the explicit CP-violation properties of the 2HDM scalar potential are
$I_{Y3Z}$ [\eq{yzzz}], $I_{2Y2Z}$
[\eq{yyzz}], $I_{6Z}$ [\eq{sixz}] and $I_{3Y3Z}$ [\eq{i3y3zform}].
We have shown that a real basis exists,
implying that the 2HDM potential is 
explicitly CP-conserving, if and only if
$I_{Y3Z}=I_{2Y2Z}=I_{6Z}=I_{3Y3Z}=0$. We refer to these
invariant imaginary parts as \iinv s.  

Note that the above conditions are not
sufficient to guarantee that the scalar sector conserves CP, since the
minimization of the scalar potential may generate complex vacuum
expectation values (vevs).  As stated above, if the vevs
possess a non-zero relative phase in \textit{all}
real basis choices, then the model spontaneously breaks CP.
One can formulate basis independent conditions for spontaneous
CP-violation.  First, one must prove that the Higgs sector is
\textit{explicitly} CP-conserving (the
corresponding invariant conditions have been given above).
Spontaneous CP-violation depends on the properties of the Higgs field vevs,
$v_a$, which can be
combined with the Higgs potential parameters to construct additional invariant
quantities.  Such invariant conditions have been previously obtained in
\Ref{lavoura}, and are exhibited in \sect{sec:five}.  Combining the
information from these two classes of invariant conditions, one can
distinguish between explicit and spontaneous CP-violation in the 
2HDM.

The phenomenological consequences of our invariants will be considered
in a forthcoming paper.  To apply the basis-independent technology to
experimental studies, one would have to examine various CP-violating
observables and express them in terms of our invariant quantities.
The LHC would provide the first possible arena for such studies.
However, the number of Higgs observables that could be extracted from
LHC analyzes is limited.  We anticipate that Higgs-mediated
CP-violating effects are likely to be small, and their extraction will
surely require precision measurements.  A future high energy
$e^+e^-$ linear collider such as the ILC could provide the required
luminosity and precision to begin a program of CP-violating Higgs
phenomenology.  We plan on examining possible CP-violating observables
and determining their sensitivity to the
\iinv s.  This analysis will require a better
understanding of the relation of the \iinv s to the mixing of
CP-even/CP-odd neutral Higgs boson eigenstates.

Perhaps the most attractive 2HDM model is the
one associated with the minimal supersymmetric extension 
of the Standard Model (MSSM)~\cite{mssm}.  
Indeed, the tree-level Higgs sector of
the MSSM is CP-conserving.  However, when loop-effects are included,
supersymmetry breaking effects, which enter via the loops, can
impart non-trivial phases to parameters of the effective 2HDM scalar
potential~\cite{Haber:1993an,cpsusy}.\footnote{These 
phases would be directly related to phases
of fundamental complex MSSM parameters such as the
supersymmetric-conserving $\mu$-term, and the
supersymmetry-breaking gaugino Majorana mass
terms and matrix $A$-parameters.}
One can therefore express the \iinv s in terms of
fundamental MSSM parameters.  This may lead to relations among the
four \iinv s introduced above, depending on the model of supersymmetry
breaking.

Ultimately, if nature employs a 2HDM as an effective theory of
electroweak symmetry breaking, it will be crucial to determine whether
Higgs-mediated CP-violation exists and determine its structure.
By devising experimental probes of the four 
\iinv s, we hope to provide a model-independent technique for
elucidating the fundamental theory that is  responsible for
Higgs sector dynamics.

\acknowledgments

We have greatly benefited from discussions with Sacha Davidson 
and her participation in the initial stages of this work.
We are also grateful for the hospitality of the Aspen Center for
Physics, where this project was initiated.

This work was supported in part by the U.S. Department of Energy.

\appendix

\section{Existence of a Real Basis}

In this appendix, we prove Theorem 1 that was quoted at the beginning
of \sect{sec:three}.

\noindent {\underline{\bf Theorem 1:}}  The Higgs potential is
explicitly
CP-conserving if and only if a basis exists in which all Higgs potential
parameters are real.  Otherwise, CP is explicitly violated.

A basis in which all Higgs potential parameters are real will be
called a \textit{real basis}.
In order to prove Theorem 1, one can either consider the most general CP
transformation laws of the scalar fields or invoke the 
CPT theorem~\cite{cpt} and
consider the most general scalar field transformation laws under
time-reversal.  Here we choose the latter procedure.\footnote{In
\Ref{branco}, the CP transformation of the scalar fields in the real
basis are used to prove that the scalar Lagrangian is CP-invariant.}
Following \Ref{Branco:1983tn}, we note that
the form for the action of the
anti-unitary time-reversal operator $\tmz$ on a set of
scalar field multiplets is given by
\beq \label{treq1}
\tmz\Phi_a(\mathbold{\vec{x}},t)\tm1= e^{i\psi}(U_T)_{a\bbar}\Phi_b
(\mathbold{\vec{x}},-t)\,,
\quad\qquad
\tmz\Phi^\dagger_\abar(\mathbold{\vec x},t)\tm1=\Phi^\dagger_\bbar
(\mathbold{\vec x},-t)(U_T^\dagger)_{b\abar}e^{-i\psi}\,.
\eeq
where $U_T$ is a \textit{symmetric} unitary matrix that depends on the
choice of basis.  The arbitrary phase factor $e^{i\psi}$ corresponds
to the freedom to make U(1)$_{\rm Y}$ transformations.\footnote{More 
generally, the time reversal operator is defined modulo 
SU(2)$\times$U(1)$_{\rm Y}$ gauge transformations that
leave the Lagrangian invariant (and hence do not modify the scalar
potential parameters).}  
To prove that
$U_T$ is symmetric, we apply the time reversal operator twice
and use the well known result that $\tmz^2\Phi_a(\mathbold{\vec x},t)
{\mathcal{T}}^{-2}=
\Phi_a(\mathbold{\vec x},t$); that is, $\tmz^2=1$ when applied to
a bosonic field~\cite{carruthers}.   Applying this result to \eq{treq1}
yields $U_T^* U_T=I$, due to the anti-unitarity of
$\tmz$.  Since $U_T$ is unitary, it follows that $U_T$ must
satisfy $U_T^T=U_T$.  The (canonical) kinetic energy terms of the
scalar field theory are automatically time-reversal invariant.  It then
follows that the scalar Lagrangian is time-reversal invariant if the
scalar potential satisfies:\footnote{We henceforth omit
exhibiting the
explicit dependence of the fields on the space-time coordinates.}
\beq \label{linvariance}
\tmz\,\mathcal{V}(\Phi,\{p\})\,\tm1=
\mathcal{V}(U_T\Phi,\{p^*\})=\mathcal{V}
(\Phi,\{p\})\,,
\eeq
where $\left\{p\right\}$ represents the
Higgs potential parameters appearing in
$\mathcal{V}$, and the complex conjugated parameters $\left\{p^*\right\}$
appear above due to the anti-unitarity of $\tmz$.  If \eq{linvariance}
is satisfied, then the action is invariant under time-reversal
transformations.

Suppose that a basis exists in which all the Higgs potential
parameters are real.  In this case, we may choose $U_T=1$, in which
case \eq{linvariance} is trivially satisfied.  To complete the proof
of Theorem 1, we must show that
a basis exists in which all the Higgs potential parameters are real
if \eq{linvariance} is satisfied.
First, we examine the quadratic part of the Higgs potential, which we
can write in matrix notation as:
\beq
\mathcal{V}_2=\Phi^\dagger Y\Phi\,,
\label{lagy}
\eeq
where $Y$ is a hermitian matrix.  Time reversal invariance of
$\mathcal{V}_2$ requires
\beq
\tmz \Phi^\dagger Y\Phi\tm1=\Phi^\dagger
U_T^\dagger Y^* U_T\Phi=\Phi^\dagger Y \Phi\,,
\label{ltransform}
\eeq
where we have used $\tmz Y\tm1=Y^*$.
\Eq{ltransform} implies that
\beq
U_T^\dagger Y^* U_T=Y\,.
\label{req1}
\eeq
As shown in Appendix~\ref{app:matrixproof},
since $U_T$ is unitary and symmetric, we can write
\beq
\label{eqv}
U_T=V^T V,
\eeq
where $V$ is unitary (but not necessarily symmetric).
As a result, \eq{req1} will be true if
\beq 
V^\dagger V^* Y^* V^T V=Y\,,
\eeq
which can be converted to
\beq \label{vyv}
 (VYV^\dagger)^*=VYV^\dagger \,.
\eeq
That is $Y'\equiv VYV^\dagger$ is real.  But, $Y'$ is simply $Y$ in
the new basis $\Phi'=V\Phi$.
Thus, there exists a basis in which the parameters
of $\mathcal{V}_2$ are real.

A similar computation can be performed for the rest of the terms
appearing in the scalar potential.  In particular, if we write the
quartic part of the Higgs potential as:
\beq
\mathcal{V}_4=\half Z_{a\bbar c\dbar}(\Phi^\dagger_\abar\Phi_b)
(\Phi^\dagger_\cbar\Phi_d)\,,
\eeq
then the analog of \eq{req1} is
\beq
(U_T^\dagger)_{e\abar}(U_T)_{b\fbar}(U^\dagger_T)_{g\cbar}(U_T)_{d\bar h}
Z_{a\bbar c\dbar}^*=Z_{e\fbar g \bar h}\,.
\eeq
We again apply \eq{eqv} and conclude that
\beq \label{vzv}
[V_{p\abar}V^\dagger_{b\bar q}V_{r\cbar}V^\dagger_{d\bar s}
Z_{a\bbar c\dbar}]^*=
V_{p\ebar}V^\dagger_{f\bar q}V_{r\gbar}V^\dagger_{h\bar s}
Z_{e\fbar g\bar h}\,.
\eeq
That is, the unitary transformation $V$ produces the
basis in which \textit{all} the Higgs potential parameters are real.

Conversely, if no basis exists in which the Higgs potential parameters
are real, then no unitary matrix $V$ exists such that \eqs{vyv}{vzv}
are simultaneously satisfied. Following the above proof in the
backward direction, one can conclude that no choice of a unitary
symmetric matrix $U_T$ exists that satisfies \eq{linvariance}.

In some cases (see below), more
than one suitable time-reversal operator exists.
Any one of these operators can be used to demonstrate
that the Higgs potential is explicitly CP-invariant.
Nevertheless, in order to ascertain that the Higgs
sector is invariant under CP, it is necessary
to verify that the vacuum is also CP-invariant 
(equivalently time-reversal invariant).   
In particular, the vacuum may select out a unique time-reversal
operator, as shown in Appendix F.  (If the vacuum is non-invariant with
respect to all possible candidate time-reversal operators, then
time-reversal invariance is spontaneously broken.)
Thus, it is important to consider the possible non-uniqueness in
the definition of $\mathcal{T}$ given in \eq{treq1}.

For an explicitly CP-conserving Higgs potential,
a real basis must exist.
However, the real basis is not unique.
In particular, given a real $\Phi'$-basis, there
exists an O(2)$\times\mathcal{D}$ subgroup of U(2)
consisting of $2\times 2$ unitary matrices
$W_{a\bbar}$ such that the scalar potential parameters
remain real under $\Phi'_a\to \Phi''_a=W_{a\bbar}\Phi'_b$.
Here, $\mathcal{D}$ is the maximal discrete subgroup of U(2)
that is a symmetry of the Higgs Lagrangian.
In addition, one is free to make U(1)$_{\rm Y}$ phase
rotations, which simply reflects the fact that $U_T$ is only defined up
to an overall phase.
If $\mathcal{D}$ is trivial, then $W$ is an orthogonal transformation
and $U_T=I$ (up to an overall phase) in \textit{any} real basis.
If $\mathcal{D}$ is nontrivial, then $W^T W\neq e^{i\eta}I$
(for any phase choice $\eta$),  in which case
the choice of $U_T$ in the definition of the time reversal operator
is not unique (modulo gauge transformations).

To amplify these remarks, we suppose that in the original $\Phi$-basis
another anti-unitary operator $\widetilde{\mathcal{T}}$
exists that is a potential candidate for the time-reversal operator.
In particular, suppose that
there exists a symmetric unitary matrix $\widetilde U_T\neq e^{i\eta}U_T$
such that\footnote{That is, if
$\widetilde{U_T}\neq U_T$ in the $\Phi$-basis, 
for any choices of the phases $\widetilde\psi$ and $\psi$, then
$\widetilde{\mathcal{T}}$ and $\mathcal{T}$ are distinct and equally
valid choices for the time reversal operator.}
\beq \label{wtiltime}
\ttmz\Phi_a(\mathbold{\vec{x}},t)\ttm1=
e^{i\widetilde\psi}(\widetilde U_T)_{a\bbar}\Phi_b
(\mathbold{\vec{x}},-t)\,,
\qquad\qquad
\mathcal{V}(\Phi,\{p\})=\mathcal{V}(\widetilde U_T\Phi,\{p^*\})\,.
\eeq
Then, the analysis above
implies that there exists a unitary matrix $\widetilde V$ such that
$\widetilde U=\widetilde V^T\widetilde V$, and $\Phi''=\widetilde
V\Phi$ is also a real basis.  In this case, the real $\Phi'$-basis and
the real $\Phi''$-basis are related by $\Phi''=W\Phi'$ where
$W=\widetilde V V^{-1}$.  It follows that  
$WW^T=\widetilde VU^{-1}\widetilde V^T\neq
e^{i\eta}I$ (for any phase choice $\eta$).
Thus, the existence of $\widetilde{\mathcal{T}}\neq\mathcal{T}$ 
implies that the discrete group $\mathcal{D}$ is nontrivial.
Likewise, one can show that $W^T W=[V^{-1}]^T\widetilde UV^{-1}\neq
e^{i\eta}I$.

Given $U_T$ in the $\Phi$-basis, we may determine the form of this
matrix in any real basis.  For example, inserting $\Phi'=V\Phi$ into
\eq{treq1} and making use of \eq{eqv}, we find
\beq
\tmz\Phi'_a(\mathbold{\vec{x}},t)\tm1= e^{i\psi}\Phi'_a
(\mathbold{\vec{x}},-t)\,.
\eeq
That is, in the $\Phi'$-basis, $U'_T=I$.  \Eq{linvariance} then implies
that this is a real basis.  Now, let us transform to the real
basis $\Phi''=W\Phi'$.  A similar computation yields
\beq \label{phipp}
\tmz\Phi''_a(\mathbold{\vec{x}},t)\tm1= e^{i\psi}(WW^T)^{-1}_{a\bbar}\Phi''_b
(\mathbold{\vec{x}},-t)\,.
\eeq
where $U''_T=(WW^T)^{-1}\neq I$ in the $\Phi''$-basis.
Similarly, if we identify $\widetilde{\mathcal{T}}$ as the time-reversal
operator, we find that
$\widetilde U'_T=W^T W\neq I$ and $\widetilde U''_T=I$.  
We may assemble all possible real bases into classes.  Each class is
in one-to-one correspondence with the elements of the discrete group
$\mathcal{D}$.  In the class of real bases associated with the
identity element of $\mathcal{D}$, the corresponding $U_T=I$.  In all
other classes of real bases, the corresponding $U_T\neq I$.

If $\mathcal{D}$ is trivial, so that $W$ is an orthogonal
transformation [up to an overall phase that can be absorbed,
\textit{e.g.}, into the
multiplicative phase factor in \eq{phipp}],
then $U_T=I$ in \textit{any} real basis.  In this case, the definition of 
the time reversal operator ${\mathcal{T}}$ is unique (modulo gauge
transformations).  

Finally, we note that the existence of a non-trivial discrete subgroup
$\mathcal{D}$ imposes strong constraints on the parameters of the
Higgs potential.  Consider a real $\Phi'$-basis and a real
$\Phi''$-basis related by $\Phi''=W\Phi'$.  It then follows that
$Y''=WY'W^\dagger$.  By assumption,
$Y'$ and $Y''$ are real.  A short computation
then yields the vanishing of the following commutators:
\beq \label{commute}
[Y',W^T W]=[Y'',WW^T]=0\,.
\eeq
A similar constraint arises from the requirement that both $Z'$ and
$Z''$ are real.  
Using these results, it is straightforward to verify that
\eq{wtiltime} is satisfied for $\widetilde U'_T=W^T W$ in the
$\Phi^\prime$-basis and  \eq{linvariance} is satisfied for
$U''_T=(WW^T)^{-1}$ in the $\Phi''$-basis.

\section{A proof of a result from matrix analysis}
\label{app:matrixproof}

In the proof of Theorems 1 and 4,
the following lemma is required:

\underline{\bf Lemma 1:}
A complex $n\times n$ matrix $U$ is unitary and symmetric if and only if
there is a complex $n\times n$ unitary matrix $V$ such that $U=V^T V$.

This result is given as problem 17 on p.~215 of \Ref{horn}.  Here, we
give an explicit proof.  Clearly if $V$ is unitary it follows that $U$
is unitary and symmetric.  Thus, we focus on the proof that given $U$,
the unitary matrix $V$ exists.  Lemma~1 is a special case of the Takagi
factorization of a complex symmetric matrix (see pp.~204--206 of \Ref{horn}).
Namely, for any
complex symmetric matrix $M$, there exists a unitary matrix $V$ such
that $M=V^TDV$, where $D$ is a real nonnegative diagonal matrix whose
elements are given by the nonnegative square roots of the eigenvalues
of $MM^\dagger$.\footnote{The Takagi
factorization of a complex symmetric matrix is the basis for the
mass diagonalization of a general Majorana fermion mass
matrix~\cite{mohapatra}.}  Applying the Takagi factorization to a
unitary matrix $M=U$ (\ie, $UU^\dagger=I$), it immediately
follows that $D=I$.  Hence, $U=V^T V$ for some unitary matrix $V$.

The matrix $V$ is not unique.  In particular, if $U=V^TV$ then
$U=W^T W$, where the
unitary matrix $W=KV$ and $K$ is an arbitrary orthogonal matrix.
However, the proof of Theorem 4 simply requires the existence of $V$,
which has been proven above.

\section{Does a basis exist in which all
the $\boldsymbol{\lambda_i}$ are real?}

{\underline{\bf Lemma 2:}} If the parameters of the 2HDM satisfy the
relations: $\lam_1=\lam_2$ and $\lam_7=-\lam_6$, then one can always
transform to a new basis in which
$\lam_5'$ and $\lam_7'=-\lam_6'$ are all real.

We begin with \eq{Lam5def} and \eq{Lam6def} and require that
the imaginary parts of $\lam_5'$ and $\lam_6'$ are
zero.  We assume that $\lam_6\neq 0$
(if $\lam_7=-\lam_6=0$, it is trivial to transform to a
basis where $\lam_5$ is real by rephasing one of the scalar fields).
Moreover, without loss of generality, we may assume that $\lambda_6$
is real by rephasing one of the scalar fields
appropriately.\footnote{Since $\lambda_1=\lambda_2$ and
$\lambda_7=-\lambda_6$, it follows that $\lambda'_7=-\lambda'_6$, and
further consideration of $\lambda_7$ is unnecessary.}
If $\lambda_5$ is also real after the rephasing, we are done.  If not,
we write $\lambda_5\equiv\alv e^{i\theta_5}$ and obtain
\beqa
{\rm Im}~\lambda'_5&=& -\half f_b\sin 2\chi+f_a\cos 2\chi\,,\label{pr5}\\
{\rm Im}~\lambda'_6&=& -\quarter f_d\sin \chi+\half f_c\cos \chi\,,
\label{pr6}
\eeqa
where
\bea
f_a&=&
     {\alv}\,c_{2\theta}\sin({\theta_5} + 2\,\xi)
-2\lambda_6 s_{2\theta}\sin\xi\,,\label{fa}\\
f_b&=&(\lam_1-\lam_3-\lam_4)s^2_{2\theta}+
 \alv (2-s^2_{2\theta})\cos ({\theta_5} + 2\,\xi) -
2\lambda_6 s_{4\theta}\cos\xi\,,\label{fb} \\
f_c&=& \alv s_{2\theta}\sin ({\theta_5} + 2\,\xi)+2\lambda_6
c_{2\theta}\sin\xi \,,\label{fc} \\
f_d&=&\left[\alv\cos({\theta_5} + 2\xi)-\lambda_1+\lambda_3+\lambda_4)
\right]s_{4\theta}+4\lambda_6 c_{4\theta}\cos\xi\,. \label{fd}
\eea
As before, we abbreviate $s_{4\theta}\equiv\sin 4\theta$,
$c_{4\theta}\equiv\cos 4\theta$, \etc\
We proceed to solve ${\rm Im}~\lambda'_5=0$,
which yields an equation for $\cot 2\chi$, and ${\rm Im}~\lambda'_6=0$,
which yields an equation for $\cot\chi$:
\bea
\cot 2\chi&=&{\frac{f_b}{2f_a}}\label{tantwochi}\\[5pt]
\cot\chi&=&{\frac{f_d}{2f_c}},\label{tanchi}
\eea
Under the assumption that $f_a\neq 0$ and $f_c\neq 0$,
we can eliminate
$\chi$ by employing the well known identity
\beq \label{cottwo}
\cot 2\chi=\frac{\cot^2\chi-1}{2\cot\chi}\,,
\eeq
which leads to the following result:
\beq \label{master}
G(\theta,\xi)\equiv f_a(f_d^2-4f_c^2)-2f_b f_c f_d=0\,.
\eeq
We wish to prove that there exists at least one
$\theta$ and $\xi$ that solves \eq{master}.
From any such solution, we may compute $\chi$ from \eqs{tantwochi}{tanchi}.
This would then provide the elements of the U(2) transformation matrix
that yields the basis in which all the $\lambda_i$ are real.

To prove that a solution to $G(\theta,\xi)=0$ exists, we note that
\bea
f_a(\theta=0,\xi)&=&-f_a(\theta=\pi/2,\xi)=\alv \sin(\theta_5+2\xi)\,,\\
f_b(\theta=0,\xi)&=&+f_b(\theta=\pi/2,\xi)=2\alv \cos(\theta_5+2\xi)\,,\\
f_c(\theta=0,\xi)&=&-f_c(\theta=\pi/2,\xi)=2\lambda_6\sin\xi\,,\\
f_d(\theta=0,\xi)&=&+f_d(\theta=\pi/2,\xi)=4\lambda_6\cos\xi\,,
\eea
from which it follows that
\beq
G(0,\xi)=-G(\pi/2,\xi)=16\lambda_6^2|\lam_5|\sin\theta_5\,.
\eeq
This means that $G$ will have
at least one sign change as a function of $\theta$.
Hence, for any value of $\xi$
there exists a value of $\theta$ for which $G(\theta,\xi)=0$.
Thus, we have proved the existence of a U(2) transformation that
results in a basis in which all the $\lambda_i$ are real.

The assumption above that $f_a\neq 0$ and $f_c \neq 0$ for values of
$\theta$ and $\xi$ at which $G(\theta,\xi)=0$ is not
strictly necessary.  For example, if $f_a=0$ (but $f_b\neq 0$), then
one can rewrite
\eq{tantwochi} in terms of $\tan 2\chi$.  We then end up again with
\eq{master}.  The only special cases that need be considered are:
(i) $f_a=f_b=0$ and (ii)~$f_c=f_d=0$.  If (i) and (ii) both hold, then
we immediately conclude that $\lambda'_5$ and $\lambda'_6$ are real
and we are finished.  If only (i) [only (ii)] holds, then we simply
use \eq{tanchi} [\eq{tantwochi}] to determine $\chi$, and we are finished.

We have used Lemma 2 in the proof of Theorem 2 (see Section III).  It
is instructive to examine the necessity of
the condition of $\lambda_1=\lambda_2$ in the proof of Lemma 2.
For this reason, we prove a second lemma.

{\underline{\bf Lemma 3:}}
If $\lam_1\neq\lam_2$ and ${\rm Im}(\lambda_5^*\lambda_6^2)\neq 0$
in a basis where $\lam_7=-\lam_6\neq 0$, then
it is impossible to transform to a basis
in which $\lam_5'$, $\lam_6'$ and $\lam_7'$ are all real.

The proof of Lemma 3 is trivial using invariants.  Namely, in a basis
where $\lambda_7=-\lambda_6\neq 0$, we
use \eq{sixzee} to conclude that $I_{6Z}\neq 0$.
Hence in this case, there is no
basis in which all the $\lambda_i$ are real.

Even without invariants, it is not difficult to show that no basis
exists in which all the $\lambda_i$ are real.  We
first rephase one of the scalar
fields such that the resulting value of $\lambda_6$ is real.
In this basis, $\lambda_5\equiv |\lambda_5|e^{i\theta_5}$, where 
$\theta_5\neq 0~(\rm{mod}~\pi)$.
We then use
\eqs{Lam6def}{Lam7def} in the case of $\lambda_7=-\lambda_6$ to obtain
\beq
\Im(\lam_6'+\lam_7')=\half \sin\chi\,\stwot(\lam_1-\lam_2)\,.
\label{im67}
\eeq
Since $\lam_1\neq\lam_2$, it follows that 
$\Im\lam_6'=\Im\lam_7'=0$ implies that either
$\sin2\theta=0$ or $\sin\chi=0$.

If $\sin 2\theta=0$, then \eqs{Lam5def}{Lam6def} yield
\beqa
\lambda'_5 e^{2i\chi}&=&|\lambda_5|e^{\pm i(2\xi+\theta)}\,,\\
\lambda'_6 e^{i\chi}&=&|\lambda_5|e^{\pm i\theta}\,,
\eeqa
where the choice of sign above corresponds to the sign of $\cos
2\theta$.  Thus,
\beq
\frac{\lambda'_5}{\lambda^{\prime\,2}_6}=\frac{|\lambda_5|}{\lambda_6^2}
e^{\pm i\theta_5}\,,
\eeq
and we see that no basis exists in which $\lambda'_5$ and $\lambda'_6$
are simultaneously real.

Next, suppose that $\sin 2\theta\neq 0$ and $\sin\chi=0$.
\Eqs{Lam5def}{Lam6def} then yield
\bea
\Im(\lam_5')&=&\alv c_{2\theta}\sin(\theta_5+2\xi)
-2\lam_6 s_{2\theta}\sin\xi\,,\label{l5pr} \\
\Im(\lam_6')&=&\half\alv s_{2\theta}\sin(\theta_5+2\xi)
+\lam_6 c_{2\theta}\sin\xi\,.\label{l6pr}
\eea
If $\sin\xi=0$, then \eq{l6pr} reduces to
$\Im(\lam_6')=\half|\lambda_5|\sin 2\theta\sin\theta_5\neq 0$.  Thus,
we must have $\sin\xi\neq 0$ if $\lambda'_6$ is real.  Then, using
\eqs{l5pr}{l6pr} to set $\Im(\lam_5')=\Im(\lam_6')=0$ yields
\beq \label{tancot}
\tan2\theta=-\cot 2\theta=
\frac{\alv\sin(\theta_5+2\xi)}{2\lam_6\sin\xi}\,.
\eeq
However, \eq{tancot} implies that $\tan^2 2\theta=-1$, which is
impossible.  Once again, we conclude that no basis exists in which
$\lambda'_5$ and $\lambda'_6$ are simultaneously real.  The proof of
Lemma 3 is now complete.

\section{Proof of Lemma 4}

Consider the special isolated point of the Higgs parameter
space in which $\lambda_1=\lambda_2$ and $\lambda_7=-\lambda_6$.  By
Lemma 2, we may assume without loss of generality that all the
$\lambda_i$ are real.  Thus, $Y_{12}$ remains as the only potentially
complex parameter. Lemma 4 provides the conditions under which it is
possible to find a new basis in which all the Higgs potential
parameters are real.

{\underline{\bf Lemma 4:}} If the parameters of the 2HDM satisfy the
relations: $\lam_1=\lam_2$ and $\lam_7=-\lam_6$, and the basis is
chosen such that all the $\lambda_i$ are real and $Y_{12}$ is complex,
then there exists a new basis in which all the Higgs potential
parameters are real if and only if (at least) one of the following
two conditions is satisfied:
\beq \label{condi}
\lam_5^2+\lam_5 (\lam_1-\lam_3-\lam_4) - 2\lam_6^2=0\,,
\eeq
and/or
\beq \label{condii}
 4\lam_6\,({\rm Re}~Y_{12})^2-
(\lam_3+\lam_4+\lam_5-\lam_1)(Y_{11}-Y_{22})\,{\rm Re}~Y_{12} -\lam_6
(Y_{11}-Y_{22})^2=0\,.
\eeq

It is easy to prove that if neither \eq{condi} nor \eq{condii} is
satisfied, then there is no basis in which all Higgs potential
parameters are real.  The latter conclusion follows directly from
$I_{3Y3Z}\neq 0$, which is a consequence of \eq{3y3zspecial}.
Thus, we focus on the inverse statement: if either
\eq{condi} or \eq{condii} is satisfied, then there exists a basis
in which all the Higgs potential parameters are real.

Suppose that \eq{condi} is satisfied, under the
assumption that all the $\lambda_i$ are real (for $\lam_1=\lam_2$
and $\lam_7=-\lam_6$) and $Y_{12}$ is complex.
We search for a U(2) transformation to a new basis in which the
$\lambda'_i$ and $Y'_{12}$ are real.  It will be
sufficient to consider solutions with $\chi=\pi/2$.
At this point, we assume that
$\lambda_6\neq 0$ (we shall treat the case of $\lambda_6=0$ separately).
Then, we demand that $\theta$ is the solution (as a function of $\xi$)
of the following equation:
\beq \label{fasol}
\lambda_6\sin 2\theta=\lambda_5\cos 2\theta\cos\xi\,.
\eeq
Using \eq{pr5} with $\chi=\pi/2$ and real $\lambda_5$, it is easy to
check that \eq{fasol} implies that
${\rm Im}~\lambda_5'=0$.
Next, using
\eq{pr6} with $\chi=\pi/2$ and real $\lambda_5$ yields:
\beq \label{fdsol}
{\rm Im}~\lambda_6'=-\quarter
(\lambda_5\cos 2\xi-\lambda_1+\lambda_3+\lambda_4)\sin 4\theta
-\lambda_6\cos 4\theta\cos\xi\,.
\eeq
Using \eq{fasol}, we obtain
\beqa
\sin 4\theta&=&2\sin 2\theta\cos 2\theta=\frac{2\lambda_5\cos^2 2\theta
\cos\xi}{\lambda_6}\,,\\
\cos 4\theta&=&\cos^2 2\theta-\sin^2 2\theta=\cos^2 2\theta\left(1-
\frac{\lambda_5^2\cos^2\xi}{\lambda_6^2}\right)\,.
\eeqa
Inserting these results into \eq{fdsol} and simplifying the resulting
expression yields
\beq
{\rm Im}~\lambda_6'=
\frac{\cos^2 2\theta\cos\xi}{2\lambda_6}\biggl[\lambda^2_5
+\lambda_5(\lambda_1-\lambda_3-\lambda_4)-2\lambda_6^2\biggr]\,.
\eeq
Thus, using \eq{condi}, we see that ${\rm Im}~\lambda_6'=0$ for
\textit{any} value of $\xi$.
We now choose $\xi$ in order that
${\rm Im}~Y_{12}'=0$.  Using \eq{mab} with $\chi=\pi/2$ and $Y_{12}\equiv
|Y_{12}|e^{i\theta_{12}}$, we find
\beq \label{yonetwozero}
2|Y_{12}|\cos 2\theta\cos(\theta_{12}+\xi)=(Y_{11}-Y_{22})\sin 2\theta\,.
\eeq
Using \eq{fasol} to eliminate $\theta$, we end up with
\beq \label{xisol}
\tan\xi=\cot\theta_{12}-\frac{\lam_5(Y_{11}-Y_{22})}{2\lam_6 |Y_{12}|
\sin\theta_{12}}\,.
\eeq

Finally, we treat the case of $\lambda_6=0$.  We may assume that
$\lambda_5\neq 0$ (otherwise, a simple rephasing of one of the Higgs
fields is sufficient to yield a real $Y_{12}$).  In this case, we
choose $\chi=\xi=\pi/2$.  Then, ${\rm Im}~\lambda'_5=0$ is satisfied
[see \eq{fasol}] for arbitrary $\theta$. Inserting $\xi=\pi/2$ into
\eq{fdsol} yields
\beq
{\rm Im}~\lambda'_6=\quarter(\lambda_5+\lambda_1-\lambda_3-\lambda_4)
\sin 4\theta=0\,,
\eeq
after using \eq{condi} with $\lambda_6=0$ and $\lambda_5\neq 0$.
We now choose $\theta$ in order that
${\rm Im}~Y_{12}'=0$.  After putting $\xi=\pi/2$ in
\eq{yonetwozero}, the end result is
\beq \label{y12alt}
\cot 2\theta=\frac{Y_{22}-Y_{11}}{2|Y_{12}|\sin\theta_{12}}\,.
\eeq

To summarize, if \eq{condi} is satisfied,
we have exhibited a U(2) transformation [\eq{u2} with $\chi=\pi/2$,
$\theta$ given by the solution to \eq{fasol} and $\xi$ given by
\eq{xisol} if $\lambda_6\neq 0$, and $\chi=\xi=\pi/2$ and $\theta$
given by the solution to \eq{y12alt} if $\lambda_6=0$]
such that all Higgs potential parameters are real in the transformed
basis.\footnote{Other U(2) transformations with $\chi\neq\pi/2$
can also produce a basis where all Higgs potential parameters are
real.  For example, a
numerical analysis suggests that if
$\chi\neq 0~({\rm mod}~\pi)$ and $\lambda_6\neq 0$,
then one can choose $\theta$ as a
function of $\xi$ such that ${\rm Im}~\lambda_5'=0$.
Using this choice for $\theta$, one again finds that
${\rm Im}~\lambda_6'=0$ as a consequence of \eq{condi}, independently
of the value of $\xi$.  Finally, $\xi$ can be chosen to yield
${\rm Im}~Y'_{12}=0$.  Of course, only one solution for $(\chi,\theta,\xi)$
must be exhibited to prove the validity of Lemma~4.}

Next, suppose that \eq{condii} is satisfied, under the
assumption that all the $\lambda_i$ are real (for $\lam_1=\lam_2$
and $\lam_7=-\lam_6$) and $Y_{12}$ is complex.
We again search for a U(2) transformation to a new basis in which the
$\lambda_i$ are still real and $Y_{12}$ is real.
In this case, we choose $\chi=\pi/2$ and $\xi=\pi$.
For this choice, $\Im\lam_5'=0$ is automatic (independently of the
value of $\theta$).  Again, we first assume that $\lambda_6\neq 0$
(the case of $\lambda_6=0$ is treated separately).  Then,
the constraints  $\Im~Y_{12}'=0$ [\eq{yonetwozero}]
and $\Im~\lam_6'=0$ [\eq{fdsol}] reduce to
\bea
\cot 2\theta&=&\frac{Y_{22}-Y_{11}}{2|Y_{12}|\cos \theta_{12}}
\,,\label{c1}\\
\cot 4\theta&=&\frac{\lam_5-\lam_1+\lam_3+\lam_4}{4\lam_6}\,,\label{c2}
\eea
respectively.  Using the double-angle formula analogous to
\eq{cottwo}, we may combine \eqs{c1}{c2} to yield the
following constraint:
\beq
4\lambda_6|Y_{12}|^2\cos^2\theta_{12}+(\lam_5-\lam_1+\lam_3+\lam_4)
(Y_{22}-Y_{11})|Y_{12}|\cos\theta_{12}-\lam_6(Y_{22}-Y_{11})^2=0\,,
\eeq
which is identical to \eq{condii}, which is assumed to be satisfied.
Thus, \eqs{c1}{c2} are consistent and provide a solution for $\theta$.

Finally, we examine the case of $\lambda_6=0$.  In this case,
\eq{condii} reduces to:
\beq
(\lam_1-\lam_3-\lam_4-\lam_5)(Y_{11}-Y_{22})\cos\theta_{12}=0\,.
\eeq
The case of $\lam_1-\lam_3-\lam_4-\lam_5=0$ (with $\lam_6=0$) is
equivalent to \eq{condi} and has already been treated.  Thus, it is
sufficient to examine only the cases of $Y_{11}=Y_{22}$ and
$\cos\theta_{12}=0$.  In both cases, we may choose $\chi=\pi/2$ and
$\xi=\pi$ as before.  Then, it is easy to check that if $\cos 2\theta=0$
in the case of $Y_{11}=Y_{22}$ and $\sin 2\theta=0$
in the case of $\cos\theta_{12}=0$, the U(2) transformation yields
${\rm Im}~\lambda_5'={\rm Im}~Y'_{12}=0$.

To summarize, if \eq{condii} is satisfied,
we have exhibited a U(2) transformation [\textit{e.g.}, \eq{u2} with 
$\chi=\pi/2$, $\xi=\pi$ and
$\theta$ given by the solution to \eq{c1} if
$\lambda_6\neq 0$]
such that all Higgs potential parameters are real in the
transformed basis.

Thus, we have explicitly constructed a U(2) transformation
that renders all Higgs potential parameters real if either \eq{condi}
or \eq{condii} is satisfied.  Consequently,
$I_{3Y3Z}=0$, and it follows that if $\lambda_1=\lambda_2$ and
$\lambda_7=-\lambda_6$, then the condition $I_{3Y3Z}=0$ is the necessary
and sufficient condition for an explicitly CP-conserving Higgs
potential.  This concludes the proof of Lemma 4.

\section{All Cubic Invariants are Real}

In this appendix, we examine invariants constructed from the
$Y_{a\bbar}$ and $Z_{a\bbar c\dbar}$.  We show that all invariants
that are at most cubic in the $Z$'s and independent of $Y$ are real.
Similarly, we demonstrate that invariants that are linear in $Y$ and
at most quadratic in the $Z$'s are real.  Finally, we prove that
invariants that are linear in $Z$ and quadratic in the $Y$'s are real.

First, we introduce some notation.  We consider all possible
non-trivial\footnote{That is,
we omit tensors that can be expressed as a product of
a scalar quantity times $Z^{(m)}_{a\bar b}$, $m=1,2$ (\textit{e.g.},
$Z_{c\bar c d\bar d} Z_{e\bar e a\bar b}\equiv
[\Tr~Z^{(2)}]Z^{(2)}_{a\bbar}$.)}
second-rank tensors that are quadratic in the $Z$'s.
Using the
symmetry properties of the $Z$'s, we find six tensors of this kind:
\beqa
&& Z^{(11)}_{c\dbar}\equiv Z^{(1)}_{a\bbar}Z_{b\abar c\dbar}\,,\qquad\qquad
 Z^{(12)}_{c\dbar}\equiv Z^{(1)}_{a\bbar}Z_{b\dbar c\abar}\,,\\
&& Z^{(21)}_{c\dbar}\equiv Z^{(2)}_{a\bbar}Z_{b\abar c\dbar}\,,\qquad\qquad
Z^{(22)}_{c\dbar}\equiv Z^{(2)}_{a\bbar}Z_{b\dbar c\abar}\,,\\
&& Z^{(31)}_{c\dbar}\equiv Z_{a\bbar e\dbar}Z_{b\abar c\ebar}\,,\qquad\qquad
\!\!
 Z^{(32)}_{c\dbar}\equiv Z_{a\bbar e\dbar}Z_{c\abar b\ebar}\,,\label{zthreej}
\eeqa
where $Z^{(1)}$ and $Z^{(2)}$ are defined in \eq{zabdef}.
A quick computation shows that the $Z^{(m)}$ ($m=1,2$) and the $Z^{(pn)}$
($p=1,2,3$ and $n=1,2$) are hermitian; that is,
\beq
Z^{(m)}_{a\bbar}=[Z^{(m)}_{b\abar}]^*\,,\qquad\qquad
Z^{(pn)}_{a\bbar}=[Z^{(pn)}_{b\abar}]^*\,.
\eeq
Next, we consider all possible non-trivial\footnote{Again,
we omit those tensors that are products of
simpler tensors (\textit{e.g.}, $Z_{a\bbar b\dbar}Z_{c\cbar e\fbar}\equiv
Z^{(1)}_{a\dbar}Z^{(2)}_{e\fbar}$).}
fourth-rank tensors that are quadratic
in the $Z$'s.    These fall into a number of
different classes.  First, we have:
\beqa
&&Z^{(1)}_{a\bbar c\dbar}\equiv Z_{a\bbar e\fbar}Z_{c\dbar f\ebar}\,,
\qquad\qquad
Z^{(4)}_{a\bbar c\dbar}\equiv Z_{a\dbar f\ebar}Z_{c\bbar e\fbar}\,,\\
&&Z^{(2)}_{a\bbar c\dbar}\equiv Z_{a\fbar e\dbar}Z_{f\bbar c\ebar}\,,
\qquad\qquad
Z^{(5)}_{a\bbar c\dbar}\equiv Z_{a\ebar f\bbar}Z_{c\fbar e\dbar}\,,\\
&&Z^{(3)}_{a\bbar c\dbar}\equiv Z_{a\fbar c\ebar}Z_{f\bbar e\dbar}\,,
\qquad\qquad
Z^{(6)}_{a\bbar c\dbar}\equiv Z_{a\fbar c\ebar}Z_{e\bbar f\dbar}\,.
\eeqa
These fourth rank tensors possess the same symmetry and hermiticity
properties as $Z_{a\bbar c\dbar}$, that is:
\beq
Z^{(n)}_{a\bbar c\dbar}=Z^{(n)}_{c\dbar a\bbar}\,,\qquad\qquad
[Z^{(n)}_{a\bbar c\dbar}]^*=Z^{(n)}_{b\abar d\cbar}\,.
\eeq
Note that $Z^{(n+3)}_{a\bbar c\dbar}\equiv Z^{(n)}_{c\bbar a\dbar}$
for $n=1,2,3$.  The second class of rank-four tensors consists of:
\beqa
\qquad\qquad
&&\widetilde Z^{(1)}_{a\bbar c\dbar}
\equiv Z_{a\bbar e\fbar}Z_{c\ebar f\dbar}\,,
\qquad\qquad
\widetilde Z^{(3)}_{a\bbar c\dbar}\equiv
Z_{a\ebar f\dbar}Z_{c\bbar e\fbar}\,,\\
&&\widetilde Z^{(2)}_{a\bbar c\dbar}\equiv
Z_{a\ebar f\bbar}Z_{c\dbar e\fbar}\,,
\qquad\qquad
\widetilde Z^{(4)}_{a\bbar c\dbar}\equiv Z_{a\dbar e\fbar}Z_{c\ebar f\bbar}\,.
\eeqa
Note that $\widetilde Z^{(n+2)}_{a\bbar c\dbar}\equiv
\widetilde Z^{(n)}_{c\bbar a\dbar}$ for $n=1,2$.
Unlike $Z_{a\bbar c\dbar}$ and $Z^{(n)}_{a\bbar c\dbar}$,
the tensors $\widetilde Z^{(n)}_{a\bbar c\dbar}$
are not symmetric under interchange of
the first and second pair of indices.  In particular,
\beq
\widetilde Z^{(1)}_{a\bbar c\dbar}=\widetilde
Z^{(2)}_{c\dbar a\bbar}\,,\qquad
\widetilde Z^{(3)}_{a\bbar c\dbar}=\widetilde
Z^{(4)}_{c\dbar a\bbar}\,.
\eeq
Consequently, we must distinguish between two types of hermiticity
conditions.  For $n=1,2$, the $\widetilde Z^{(n)}_{a\bbar c\dbar}$
satisfy the hermiticity condition of the first kind:
\beq \label{hermiticitya}
[\widetilde Z^{(n)}_{a\bbar c\dbar}]^*=
\widetilde Z^{(n)}_{b\abar d\cbar}\,,\qquad n=1,2\,,
\eeq
whereas for $n=3,4$, the $\widetilde Z^{(n)}_{a\bbar c\dbar}$
satisfy the hermiticity condition of the second kind:
\beq \label{hermiticityb}
[\widetilde Z^{(n)}_{a\bbar c\dbar}]^*=
\widetilde Z^{(n)}_{d\cbar b\abar}\,,\qquad n=3,4\,.
\eeq

The final class of rank-four tensors involve $Z^{(n)}_{a\bbar}$
($n=1,2$).  These are:
\beqa
&& Z^{(1n)}_{a\bbar c\dbar}\equiv Z_{a\bbar c\fbar}Z^{(n)}_{f\dbar}\,,
\qquad\qquad
Z^{(5n)}_{a\bbar c\dbar}\equiv Z_{a\bbar f\dbar}Z^{(n)}_{c\fbar}\,,\\
&& Z^{(2n)}_{a\bbar c\dbar}\equiv Z_{c\bbar a\fbar}Z^{(n)}_{f\dbar}\,,
\qquad\qquad
Z^{(6n)}_{a\bbar c\dbar}\equiv Z_{c\bbar f\dbar}Z^{(n)}_{a\fbar}\,,\\
&& Z^{(3n)}_{a\bbar c\dbar}\equiv Z_{c\dbar a\fbar}Z^{(n)}_{f\bbar}\,,
\qquad\qquad
Z^{(7n)}_{a\bbar c\dbar}\equiv Z_{c\dbar f\bbar}Z^{(n)}_{a\fbar}\,,\\
&& Z^{(4n)}_{a\bbar c\dbar}\equiv Z_{a\dbar c\fbar}Z^{(n)}_{f\bbar}\,,
\qquad\qquad
Z^{(8n)}_{a\bbar c\dbar}\equiv Z_{a\dbar f\bbar}Z^{(n)}_{c\fbar}\,.
\eeqa
These tensors possess neither the hermiticity nor the symmetry
properties of $Z_{a\bbar c\dbar}$.  Instead, we have (for $n=1,2$):
\beq \label{zmnstar}
[Z^{(mn)}_{a\bbar c\dbar}]^*=Z^{(m+4,n)}_{b\abar d\cbar}\,,\qquad
m=1,\ldots 4\,.
\eeq
Note that $Z^{(2n)}_{a\bbar c\dbar}\equiv Z^{(1n)}_{c\bbar a\dbar}$,
$Z^{(3n)}_{a\bbar c\dbar}\equiv Z^{(1n)}_{c\dbar a\bbar}$, and
$Z^{(4n)}_{a\bbar c\dbar}\equiv Z^{(1n)}_{a\dbar c\bbar}$ are all distinct
due to the lack of symmetry under the interchange of indices.

We proceed to examine all possible quadratic and cubic scalar
$Z$-invariants.  The quadratic scalar
$Z$-invariants are obtained by summing over the indices of the tensors
defined above in all possible allowed ways.  However, note that the
two-index tensors are hermitian, and any four-index tensor summed over
two indices yields a two-index hermitian tensor.  Hence any quadratic
$Z$-invariant is the trace of an hermitian tensor and is hence real.  We
thus turn to the (non-trivial) cubic $Z$-invariants.  These must be of
the form $Z^{(n)}_{a\bbar}X_{b\abar}$ ($n=1$ or 2) where $X_{b\abar}$
is one of the quadratic second-rank tensors defined above, or of the form
$Z_{a\bbar c\dbar}X_{b\abar d\cbar}$, where $X_{b\abar d\cbar}$ is one
of the quadratic fourth-rank tensors defined above.  But, for any
hermitian second-rank tensor, $X_{b\abar}$, the quantity
$Z^{(n)}_{a\bbar}X_{b\abar}$ is real.  Similarly, for any
fourth-rank tensor $X_{b\abar d\cbar}$ that either satisfies the
hermiticity conditions of the first or second kind
[see \eqs{hermiticitya}{hermiticityb}],
the quantity $Z_{a\bbar c\dbar}X_{b\abar d\cbar}$ is real.
All that remains is to check that the scalar quantities of the form
$Z_{a\bbar c\dbar} Z^{(mn)}_{b\abar d\cbar}$ are real.  This is proved
by first establishing the following non-trivial result:
\beq \label{specialrelations}
Z_{a\bbar c\dbar}Z^{(mn)}_{b\abar d\cbar}=Z_{a\bbar c\dbar}
Z^{(m+4,n)}_{b\abar d\cbar}\,.
\eeq
We have checked this result explicitly with Mathematica (although a simple
analytic proof eludes us).  Using \eq{zmnstar}, it immediately follows
that all such $Z$-invariants are real.
This completes the proof that all cubic $Z$-invariants are real.

We next turn to the scalar invariants that are linear in $Y$.  Since
for any hermitian two-index tensor $X_{b\abar}$, the quantity
$Y_{a\bbar}X_{b\abar}$ is real, it immediately follows that any scalar
invariant that is linear in $Y$ and at most quadratic in the $Z$'s is
real.  Finally, consider scalar invariants that are quadratic in the
$Y$'s.  Note that $Y_{a\bbar}Y_{c\dbar}$ has the same hermiticity
property as $Z_{a\bbar c\dbar}$, and $Y_{a\cbar}Y_{c\bbar}$ is an
hermitian two-index tensor.  Thus, any scalar invariant quadratic in
the $Y$'s and linear in $Z$ is real.
Hence, we have proven that all cubic invariants are real.

It is instructive to see where the above arguments break down when
quartic invariants are considered.
The simplest complex scalar invariant that is linear in
$Y$ is at least cubic in $Z$.
Indeed,
\beq
I_{Y3Z}=\Im(Z^{(1)}_{a\cbar}Z^{(11)}_{c\dbar}Y_{d\abar})=
\Im[\Tr(Z^{(1)}Z^{(11)}Y)]
\eeq
is a potentially complex quartic invariant.
Note that although $Y$, $Z^{(1)}$ and $Z^{(11)}$ are all hermitian
$2\times 2$ matrices, $I_{Y3Z}$ is not necessarily real because $Z^{(1)}$ and
$Z^{(11)}$ do not commute.   More generally, one can
check that all manifestly complex scalar invariants that are
linear in $Y$ and cubic in $Z$ can be written in the form
$\Tr(Z^{(n)}Z^{(pq)}Y)$ or $\Tr(Z^{(pq)}Z^{(n)}Y)$.
A simple Mathematica computation reveals that
\beq \label{yzzzgen}
I_{Y3Z}=\Im(Z^{(n)}_{a\cbar}Z^{(pq)}_{c\dbar}Y_{d\abar})=
-\Im(Z^{(pq)}_{c\dbar}Z^{(n)}_{a\cbar}Y_{d\abar})
\eeq
for all possible values of $n,q=1,2$ and $p=1,2,3$.  The last equality
in \eq{yzzzgen} follows from the hermiticity of the $Z^{(n)}$, $Z^{(pq)}$
and $Y$.  Hence, we conclude
that the imaginary parts of all complex invariants of this
type are equal to $\pm I_{Y3Z}$.

The simplest complex scalar invariant that is quadratic in
$Y$ is at least quadratic in $Z$.
Indeed,
\beq
I_{2Y2Z}=\Im(Y_{a\bbar}Y_{c\dbar}Z^{(11)}_{b\abar d\cbar})
\eeq
is a potentially complex quartic invariant.
This quantity is not necessarily real since $Z^{(11)}_{b\abar d\cbar}$
does not satisfy any hermiticity conditions.  More generally, one can
check that all manifestly complex scalar invariants that are
quadratic in both $Y$ and $Z$ can be written in the form
$Y_{a\bbar}Y_{c\dbar}Z^{(mn)}_{b\abar d\cbar}$ (for $m=1,\ldots,8$
and $n=1,2$).\footnote{In particular, it is straightforward to show
that $Y_{a\bbar}Y_{c\dbar}Z^{(n)}_{b\abar d\cbar}$ and
$Y_{a\bbar}Y_{c\dbar}\widetilde Z^{(n)}_{b\abar d\cbar}$ are real due
to the hermiticity properties of $Y$, $Z^{(n)}$ and $\widetilde
Z^{(n)}$.}
A simple Mathematica computation reveals that
\beq
I_{2Y2Z}=\Im(Y_{a\bbar}Y_{c\dbar}Z^{(mn)}_{b\abar d\cbar})
=-\Im(Y_{a\bbar}Y_{c\dbar}Z^{(m+4,n)}_{b\abar d\cbar})
\eeq
for all possible values of $m=1,\ldots,4$ and $n=1,2$ [where the
second equality above is a consequence of \eq{zmnstar}].
Hence, we conclude
that the imaginary parts of all complex invariants of this
type are equal to $\pm I_{2Y2Z}$.

Finally, a comprehensive analytic study of
$n$th-order pure $Z$-invariants for $n\geq 4$ of the type employed above
(in the analysis of the cubic invariants) seems prohibitive.
Thus, a systematic Mathematica-aided study was carried out
to prove that
all fourth and fifth-order $Z$-invariants are real.

\section{Time-reversal invariance of the Higgs vacuum}

In this appendix, we assume that the Higgs scalar action 
is explicitly CP-conserving
(and hence time-reversal invariant by the CPT theorem).  
That is, there exists a time reversal operator $\mathcal{T}$ that
satisfies \eq{linvariance} (for some choice of $U_T$).
In this context, we ask whether the Higgs
\textit{vacuum} is time-reversal invariant. However, there is an
apparent
ambiguity, since as shown in Appendix~A there may be a number of
distinct choices for the time reversal operator (under which the action
is invariant).  This ambiguity corresponds to a non-trivial discrete
group $\mathcal{D}$ that is a symmetry of the scalar Lagrangian.
In general, the vacuum is \textit{not} invariant with respect to 
$\mathcal{D}$.  In this case, the vacuum may select one distinct
choice for the time reversal operator.  We shall denote this
choice below by $\mathcal{T}$.   That is, 
the \textit{theory} is time-reversal
invariant if the Higgs scalar action is CP-conserving and 
the vacuum is invariant with respect to (at least) one of the
distinct choices for the time reversal operator.    
If there is no choice for the time reversal operator such
that the vacuum is invariant, then time reversal invariance
is spontaneously broken.

We denote the vacuum state by $\ket{0}$ and define
$\Phi_a\ket{0}\equiv\ket{\Phi}$.  The action of the time reversal
operator is denoted by:
\beq
\mathcal{T}\ket{0}\equiv \ket{0_T}\,,\qquad\qquad
\mathcal{T}\ket{\Phi}\equiv \ket{\Phi_T}\,.
\eeq
The anti-unitarity of $\mathcal{T}$ implies that
$\vev{0_T|\Phi_T}=\vev{0|\Phi}^*$.
Invariance of the vacuum under time-reversal invariance implies that
$\ket{0}=\ket{0_T}$.  Hence $\vev{0|\Phi_T}=\vev{0|\Phi}^*$.
It then follows that:
\beq
\vev{0|\tmz \Phi_a\tm1|0}=\vev{0|\Phi_a|0}^*\,,
\eeq
after inserting $\tmz\tm1$ in the appropriate spot
and using $\tmz\ket{0}=\ket{0}$.
Using \eq{treq1}, we end up with~\cite{Branco:1983tn}:
\beq \label{utrev}
(U_T)_{a\bbar}\vev{\Phi_b}=\vev{\Phi_a}^*\,,
\eeq
where $\vev{\Phi_a}\equiv\vev{0|\Phi_a|0}$.
We can use the above results to prove Theorem 3 of \sect{sec:five}.

\noindent {\underline{\bf Theorem 3:}}
Given an explicitly CP-conserving Higgs potential,
the vacuum is time-reversal invariant if and only if
a real basis exists in which
the Higgs vacuum expectation values are real.

We prove this theorem by demonstrating that
\eq{utrev} provides the real basis in which the vacuum
expectation values are real.  By assumption, \eq{utrev} is
satisfied in the $\Phi$-basis (which may or may not be a real basis).
As shown in Appendix~B,
one can always write $U_T=V^T V$, where the unitary matrix $V$ is
unique up to multiplication on the left by an arbitrary orthogonal
matrix.  Inserting this result into \eq{utrev} yields
\beq
V\vev{\Phi}=[V\vev{\Phi}]^*\,,
\eeq
which implies that the vacuum expectation values are real in the
$\Phi'$-basis, where $\Phi'\equiv V\Phi$.  
However, \eqs{vyv}{vzv} imply that the $\Phi'$-basis is a real
basis. Of course, if the vacuum expectation values are
real in a basis in which all the Higgs potential parameters are real,
then the choice $U_T=I$ in \eq{treq1} yields a viable time-reversal
operator. Conversely, if the Higgs scalar action is time-reversal invariant but
no real basis exists in which the vacuum expectation values are real,
then no viable time reversal transformation law exists.
In particular, no choice of $U_T$ exists that
satisfies \eq{utrev}.  This can only imply that
$\mathcal{T}\ket{0}\neq\ket{0}$.  In this case, the time reversal
symmetry is spontaneously broken.
Thus, Theorem 3 is proven.

The conditions for a time-reversal invariant theory can therefore be
reformulated.  The scalar sector of the theory is time-reversal
invariant if a $U_T$ exists that satisfies \eqs{linvariance}{utrev}.
In practice, the existence or non-existence of such a $U_T$ may be
difficult to discern, whereas the corresponding basis-independent
conditions quoted in \sect{sec:five} are straightforward to
implement.

Note that the existence of real bases does not necessarily imply that
the vacuum expectation values are real in \textit{all} possible
real basis choices.
In Appendix~A, we demonstrated that if the scalar action is
time-reversal invariant then
different choices for $\mathcal{T}$ correspond to different real bases
in which
$U_T=I$.  If the time reversal operator
is defined according to \eq{treq1} then
$U_T=V^T V$ yields a real basis $\Phi'=V\Phi$ in which $U'_T=I$.
Alternatively, if the time reversal operator is defined
according to \eq{wtiltime}, then $\widetilde
U_T=\widetilde V^T \widetilde V$ yields a real basis $\Phi''=\widetilde
V\Phi$ in which $\widetilde U''_T=I$.   
The transformation between these two real
bases is $\Phi''=W\Phi'$, where $W$ spans an O(2)$\times \mathcal{D}$
subgroup of U(2).\footnote{One can also perform a U(1)$_{\rm Y}$
transformation, which does not modify the relative phase of the two
vacuum expectation values.}
In Appendix A, we noted that $U''_T=(WW^T)^{-1}$ and $\widetilde U'_T=
W^TW$.  If $\mathcal{D}$ is trivial, then $WW^T=W^TW=I$
and $U_T=I$ (up to an overall phase) in
\textit{any} real basis.  \Eq{utrev}
then implies that the vacuum expectation values are
relatively real in any real basis (and can be chosen real with an
appropriate U(1)$_{\rm Y}$ phase rotation).  If $\mathcal{D}$ is
nontrivial, then the vacuum expectation values cannot be relatively real
in both the $\Phi'$-basis and the $\Phi''$-basis if 
$\Phi''=W\Phi'$, where $WW^T\neq e^{i\eta}I$.

As a simple example, consider again the model specified by \eq{cpc} with
$\lambda_6$ real, which was examined at the end of \sect{sec:five}.
The $\Phi$-basis in this case is a real basis but 
the vacuum expectation values,
$\sqrt{2}\,\widehat v=(e^{-i\xi/2}\,,\,e^{i\xi/2})$, exhibit a
nontrivial relative phase for $\xi\neq 0$~(mod~$\pi$).  Nevertheless,
the Higgs vacuum is time-reversal invariant.  In this case, we can
explicitly exhibit the matrix $U_T$ that satisfies \eq{utrev} and a
unitary matrix $V$ such that $U_T=V^T V$:
\beq
U_T=\left(\begin{array}{cc} 0&\quad 1\\ 1 &\quad
    0\end{array}\right)\,,\qquad\qquad
V=\frac{1}{\sqrt{2}}
\left(\begin{array}{cc}\cos\theta &\quad -\sin\theta \\
\sin\theta & \quad \phm\cos\theta\end{array}\right)\left(\begin{array}{cc}
\phm 1 &\quad  1\\ -i &\quad i\end{array}\right)\,,
\eeq
where $\theta$ is an arbitrary angle.
Indeed, the matrix $V$ transforms the (real) $\Phi$-basis to another
real basis in which the vacuum expectation values are real. In
particular, the choice of
$\theta=\xi/2$ yields $\widehat v^{\,\prime}=V\widehat v=(1,0)$ as noted at the
end of \sect{sec:five}.


\end{document}